\documentclass[journal]{IEEEtran}%
\IEEEoverridecommandlockouts
\usepackage{cite}
\usepackage{amsmath,amssymb,amsfonts}
\usepackage{amsthm,bm}
\usepackage{mathrsfs}
\usepackage{colortbl}
\usepackage{hhline}
\usepackage{graphicx}
\usepackage{subfigure} 
\usepackage{textcomp}
\usepackage{stfloats}
\usepackage{pifont}

\usepackage{enumerate}
\usepackage{booktabs}
\usepackage{url}
\usepackage{mathtools}
\usepackage{cuted}
\usepackage{array}  
\newcommand{\tabitem}{~~\llap{\textbullet}~~}
\usepackage {threeparttable}
\usepackage{tablefootnote}
\usepackage{multirow}
\usepackage{colortbl}
\def\BibTeX{{\rm B\kern-.05em{\sc i\kern-.025em b}\kern-.08em
    T\kern-.1667em\lower.7ex\hbox{E}\kern-.125emX}}

\usepackage{stackengine}
\usepackage{enumerate}
\usepackage{booktabs}
\usepackage{subfig}
\usepackage{multicol}
\usepackage[noend]{algpseudocode}
\usepackage[table,xcdraw]{xcolor}
\usepackage{pgfplots}
\usepackage{pgfplotstable}
\usepackage[ruled,linesnumbered]{algorithm2e} 
\usepackage{amssymb,amsmath,caption}
\usepackage[hang,flushmargin]{footmisc}
\usepackage{lipsum}
\makeatletter
\newcommand{\algorithmfootnote}[2][\footnotesize]{%
  \let\old@algocf@finish\@algocf@finish
  \def\@algocf@finish{\old@algocf@finish
    \leavevmode\rlap{\begin{minipage}{\linewidth}
    #1#2
    \end{minipage}}%
  }%
}
\SetKwRepeat{Do}{do}{while}
\definecolor{Reds}{RGB}{0,0,0}
\usepackage{setspace}
\usepackage{color}
\usepackage{tabularray}
\usepackage{pifont} 

\definecolor{Iron}{rgb}{0.811,0.815,0.815}
\makeatletter
\def\BState{\State\hskip-\ALG@thistlm}
\makeatother

\algnewcommand\algorithmicforeach{\textbf{for each}}
\algdef{S}[FOR]{ForEach}[1]{\algorithmicforeach\ #1\ \algorithmicdo}
\begin{document}
\title{ Agent-driven Generative Semantic Communication with Cross-Modality and Prediction

}
\author{Wanting Yang, Zehui Xiong, \textit{Senior Member, IEEE}, Yanli Yuan, Wenchao Jiang, \\ Tony Q.S. Quek, \textit{Fellow, IEEE}, Merouane Debbah, \textit{Fellow, IEEE}\thanks{

Wanting Yang, Zehui Xiong, Wenchao Jiang, and Tony Q. S. Quek are with the Pillar
of Information Systems Technology and Design, Singapore University of
Technology and Design, Singapore (e-mail: wanting\_yang@sutd.edu.sg; zehui\_xiong@sutd.edu.sg; wenchao\_jiang@sutd.edu.sg; tonyquek@sutd.edu.sg); 
Yanli Yuan is with School of Cyberspace Science and Technology, Beijing
Institute of Technology (e-mail: yanliyuan@bit.edu.cn); Merouane Debbah is with the KU 6G Research Center, Khalifa University
 of Science and Technology, Abu Dhabi, United Arab Emirates (e-mail: merouane.debbah@ku.ac.ae). 
}
}

\makeatletter
\setlength{\@fptop}{0pt}
\makeatother

\maketitle

\vspace{-1.8cm}
\begin{abstract}
In the era of 6G, with compelling visions of intelligent transportation systems and digital twins, remote  surveillance is poised to become a ubiquitous  practice. Substantial data volume and frequent updates present challenges in wireless networks. To address these challenges, we propose a novel agent-driven generative semantic communication (A-GSC) framework based on reinforcement learning. In contrast to the existing research on semantic communication (SemCom), which mainly focuses  on either semantic extraction or semantic sampling, we seamlessly integrate both by jointly considering the intrinsic attributes of source information and the contextual information regarding the task. Notably, the introduction of  generative artificial intelligence (GAI) enables the independent design of semantic encoders and decoders. In this work, we develop an agent-assisted semantic encoder with cross-modality capability, which can track the  semantic changes, channel condition, to perform adaptive semantic extraction and sampling. Accordingly, we design a semantic decoder with both predictive and generative capabilities, consisting of two tailored modules. Moreover, the effectiveness of the designed models has been verified using the UA-DETRAC dataset, demonstrating the performance gains of the overall A-GSC framework in both energy saving and reconstruction accuracy.
\end{abstract}


\newtheorem{definition}{Definition}
\newtheorem{lemma}{Proposition}
\newtheorem{theorem}{Theorem}

\newtheorem{property}{Property}

\vspace{0cm}

\section{Introduction}
As human society advances towards remote management, coupled with emerging 6G applications  like intelligent transportation system, digital twins and metaverses, remote  surveillance is poised to become a ubiquitous  practice~\cite{iyer2023survey}. The substantial data volume and frequent updates emphasize the challenges of redundancy and rigidity of information representation and transmission in   conventional  communication. Fortunately, 6G ushers in a new era of semantic communication (SemCom)~\cite{zhang2022semantic}, paving the way for more efficient and streamlined transmission.


In  literature, two prominent and non-overlapping branches of semantic communication research for  remote surveillance have emerged: deep learning (DL)-based semantic compression and age of information (AoI)-guided semantic sampling.
DL-based SemCom concentrates on each individual data unit, performing task-agnostic semantic compression first and then decompression~\cite{qin2021semantic}. Meanwhile, AoI-guided SemCom prioritizes information freshness, filtering out outdated and irrelevant information, which some literature regards as a form of compression in the \textit{temporal} domain~\cite{9380899}. To succinctly distinguish these two compression approaches operating in different domains, hereinafter, we refer to the latter  as semantic sampling and the former  as semantic extraction.

Both branches of research have demonstrated satisfactory performance and improvements in energy efficiency given constrained resources. Nonetheless,
it is crucial to emphasize that these methods merely consider the intrinsic attributes of source information such as compressibility of high-dimensional information and information freshness, which implies that most SemCom research in remote surveillance primarily focuses on the \textit{semantic} level\footnote{Shannon and Weaver suggest 
that SemCom encompasses two distinct levels~\cite{shannon1948mathematical}: semantic level, which focuses on the meaning of the transmitted symbols and effectiveness level, which is concerned with the final performance achieved by communication tasks.}.
In fact, contextual information regarding a specific task tends to play a more crucial role in reducing transmission redundancy~\cite{pappas2021goal}. Specifically,  only goal-related semantic information is necessary for the destination. Taking  traffic monitoring for instance, transmitting building details serves no purpose. Similarly, in scenarios like disaster monitoring, the value of intrinsic fresh information may be limited, while changing or unusual information holds greater benefits. Overall, integrating contextual information into the system design empowers SemCom to achieve a heightened level of \textit{effectiveness} with higher compression efficiency. 
To the best of our knowledge, there is a notable absence of research  on effectiveness-level SemCom for surveillance scenarios.

In response to the identified research gap, we propose a novel modular agent-driven generative SemCom (A-GSC) framework, which envisions a holistic redesign of information processing, sampling, transmission, and reconstruction processes. 
By incorporating intrinsic and extrinsic information, we achieve a remarkably high level of compression efficiency by seamlessly cascading semantic extraction and semantic sampling. This integration produces a synergistic effect greater than the sum of its parts.
An preliminary study has conceived a semantic change driven generative SemCom (SCGSC) framework for the remote image update system~\cite{yang2023semantic}\footnote{The preliminary work has been accepted by IEEE WCNC 2024.}. Therein, the introduction of generative artificial intelligence (GAI) transforms the semantic encoding into the determination of the PROMPT. The remarkable explainability of PROMPT allows for independent design and optimization of semantic encoders and decoders.   Specifically, in SCGSC framework, the semantic encoder first performs target segmentation to acquire the relevant semantic information called semantic map, then collectively considers the \textit{semantic change} and the \textit{AoI} to evaluate the value of information (VoI), and finally completes the semantic sampling based on  thresholds. Correspondingly, the semantic decoder takes the semantic map and static background of the remote scene as inputs to reconstruct the scene based on the diffusion model.

Building on the preliminary framework in~\cite{yang2023semantic}, this work focuses on a more realistic scenario. To characterize richer semantic information with less data, we introduce cross-modality capabilities into the SemCom design. Additionally, to present the illusion of real-time updates of remote scenes, the enhanced A-GSC framework integrates prediction capabilities, allowing the destination to continuously display remote scenes even during moments without sampling. Three key improvements distinguish this work from the SCGSC framework, as outlined below:
\begin{itemize}
\item  Instead of using semantic map, the semantic information (i.e., layout information) of scenes is transmitted in the form of text with a lower data volume. In addition to the vehicle's location information, which is characterized by the bounding box, the vehicle type information has also been integrated into the semantic information.
    \item Different from the existing VoI based semantic sampling strategies, we propose that the semantic sampling is performed by a reinforcement learning (RL) based agent. This agent dynamically refines the semantic sampling strategy in response to temporal changes of the source data, channel conditions, and performance in terms of semantic accuracy and energy consumption.
    \item 
To present the illusion of real-time updates, we integrate a predictive frame interpolation module at the destination, enabling a low-complexity real-time  prediction based on historical information and the sampling intervals, when the updated data from the source is unavailable.
\end{itemize}
The specific contributions of this work are outlined as follows.
\begin{itemize}
    \item We develop a modular semantic encoder featuring the cross modality of semantic extraction and agent-assisted semantic sampling. To guide agent learning, the Value of Information (VoI)—determined by both energy consumption and scene reconstruction semantic accuracy—is treated as the reward. Recognizing the non-linear relationship between prediction deviation and sampling interval, and aiming to mitigate the impact of the convergence process and the communication overhead caused by feedback rewards, we propose an offline \textit{stochastic} policy named Knowledge-Integrated Soft Actor Critic (K-SAC).
    \item We design a semantic decoder with both predictive and generative capabilities. To address the challenges in existing layout diffusion models, we first transform the received textual semantic information into  visual layout. Additionally, to reverse semantic encoding, we develop a diffusion-based generative
semantic inference module controlled by both the visual layout and the local static scene. This design ensures the fidelity of semantic information, while closely resembling a real remote scene. 
 Moreover, to facilitate semantic sampling with variable time intervals, we design a customized predictive frame interpolation module. 
    \item We conduct training and testing for the predictive frame interpolation and semantic inference modules based on data from UA-DETRAC~\cite{wen2020ua}. Once trained, these models function as components of the environment in the RL paradigm, connected to the source through the semantic sampling agent. The offline policy is trained using virtual experiences generated from a realistic $\mathcal{F}$ composite channel fading model~\cite{8638956}. The adaptability of the well-trained policy is demonstrated across multiple scenarios. Finally, we showcase the promising gains in energy savings and reconstruction accuracy achieved by the A-GSC framework and {\color{Reds} analyze the complexity of training and deployment the framework}.

\end{itemize}
Additionally, it should be noted that, while acknowledging the challenges of  pixel-perfect replication in generative SemCom,  such difficulties do not diminish its broad applicability. In many  scenarios, the primary concern is the preservation of essential semantic information.  For example, in  parking space surveillance, the system remains indifferent to particulars like the color of the vehicle and the dynamic background. Instead, it exclusively focuses on the vehicle location.

The rest of this paper is organized as follows. In Section~\ref{sec:related work}, we review the state of cutting-edge research in SemCom from three main  branches. Then, in Section~\ref{sec:systemovervive}, we briefly introduce the system model in the considered scenario and outline the overview of the proposed A-GSC framework. The key components in the  proposed A-GSC framework are presented in Section~\ref{sec:compression} and Section~\ref{sec:sampling}. An experimental evaluation is presented in Section~\ref{sec:simu}, followed by the conclusions of the study and future works in Section~\ref{sec:conclusion}. {\color{Reds}The main symbols have been summarized in Table~\ref{tbl: symbols}.}

\begin{table}[]
\footnotesize
 \centering
 \caption{\color{Reds}Summary of Main Notations}
\renewcommand{\arraystretch}{1.1}
\begin{tabular}{m{0.8cm}|m{7.3cm}}
\hline
\multicolumn{2}{c}{\textbf{System Design-Related Symbols}}                                                                                                          \\ \hline \hline
$t$                         & Index of sensing time interval \\ \hline
{${\bf{s}} _t$}             & {Semantic information for the monitored scene at STI $t$}                                        \\ \hline
{${\bf{\tilde s}} _t$}      & {Visualized semantic information for STI $t$ received}                        \\ \hline
{${\bf{\tilde s}}' _t$}     & {Predicted visualized semantic information for STI $t$ }                       \\ \hline
{$L _t$}                    & {Data size for the semantic information at STI $t$}                                              \\ \hline
{${\tilde g}$ (${\bar g}$)} & {Instantaneous (average) channel gain}                                                           \\ \hline
{${\bf{r}}$}                & {Static scene information of the remote scene}                                                   \\ \hline
{${\chi _t}$}               & {Semantic change degree for the monitored scene at STI $t$}                                      \\ \hline
{$P$}                       & {Number of predicted future semantic messages per round}                                         \\ \hline
{${D _t}$}                  & {Prediction deviation $D_t$ between ${{\bf{\tilde s}}_t}$ and ${{\bf{\tilde s}}' _t}$}           \\ \hline
$\hat t$      & Timestamp corresponding to the last sampled scence \\ \hline
{${\hat D_t}$}              & {Penalized prediction deviation $D_t$ between ${{\bf{\tilde s}}_t}$ and ${{\bf{\tilde s}}' _t}$} \\ \hline
{${\Delta _t}$}             & {Time interval between STI $t$ and the last semantic sample}                         \\ \hline
{${E _t}$}                  & {Energy consumption of the transmission for {${\bf{s}} _t$} at STI $t$}                             \\ \hline \hline 
\multicolumn{2}{c}{\textbf{Diffusion Model-Related Parameters}}                                                                                                                \\\hline \hline 
$n$                                & Index of denoising step \\ \hline
${{\mathbf{x}}_n}$                              & Noisy data at denoising  step $n$                                                                                    \\ \hline 
${\epsilon _n}$                                 & Noise  sampled from the standard Gaussian distribution $\mathcal{N}\left( {0,{\mathbf{I}}} \right)$  \\\hline 
${\beta _n}$                                    & Scale of added noise at each step $n$   \\ \hline   ${\bar \alpha _n}$                              & ${\bar \alpha _n}: = \prod\nolimits_{s = 1}^n {\left( {1 - {\beta _s}} \right)}$                                    \\ \hline
$\theta$                                        & Parameters of U-Net model for noise estimation                                                                      \\ \hline   \hline
\multicolumn{2}{c}{\textbf{K-SAC-Related Parameters}}                                                                                                                \\\hline \hline 
${{\bf{S}}_t}$                                  & ${{\mathbf{S}}_t} = \left[ { {L_t}, {\bm\chi _t}, {{\bar g}_t}} \right]$, the state observed by the agent at STI $t$                            \\ \hline
$W$   & Duration of observation window in units of STIs \\ \hline
${a_t}$                                         & ${a_t} \in \mathcal A = \left\{ {0,1} \right\}$, action performed by the agent at STI $t$                                                      \\ \hline
$r_t^1$                                         & Reward for performing semantic sampling                                                                                                         \\ \hline
$r_t^0$                                         & Reward for not performing semantic sampling                                                                                                   \\ \hline
$r_t$                                           & $r_t ={r_t^{\text 1}}{{\mathbf{1}}_{{a_t} = 1}} + {r_t^{\text 0}}\left( {1 - {{\mathbf{1}}_{{a_t} = 1}}} \right)$, reward obtained at STI $t$   \\ \hline
$\varphi$                                       & Parameters for critic networks                                                                                                                  \\ \hline
$\phi $                                         & Parameters for actor network                                                                                                                   \\ \hline
$\vartheta$                                     & Temperature parameter                                                                                                                           \\ \hline
$\mathcal{\bar H}$                               & Target entropy     \\ \hline                                                                    
\end{tabular}
\label{tbl: symbols}
\end{table}

\begin{table*}
\footnotesize
 \centering
 \caption{\color{Reds}Comparison of Mainstream SemCom and Proposed SemCom Frameworks.}
 \renewcommand{\arraystretch}{1.2} 
 \begin{tabular}{|p{5.6cm}|p{5.6cm}|p{5.6cm}|}
  \hline
 \multicolumn{1}{|c|}{\textbf{DL-based SemCom}}  & \multicolumn{1}{c|}{\textbf{AoI/VoI guided SemCom}} & \multicolumn{1}{c|}{\textbf{GAI-based SemCom}} \\ \hline
{\quad} \newline \vspace{ -0.6cm}\newline \textit{\small Achieved Capabilities:} \newline \tabitem Semantic extraction \& Inference {\quad} 
{\color{Reds}\newline \textit{\small Advantages:} \newline \tabitem Enhancement on the robustness for short range wireless transmission}
\newline \textit{\small Limitations:} \newline \tabitem Non-explainability \newline \tabitem Error floor in  DL
\newline \tabitem Short-range communication without relays {\quad} \newline \textit{\small Representative Refs:} \cite{qin2021semantic, zhang2023predictive,wang2022wireless, zhang2022deep,zhang2023deepma}  & {\quad} \newline \vspace{ -0.6cm}\newline \textit{\small Achieved Capabilities:} \newline \tabitem Semantic Sampling {\quad} 
{\color{Reds}\newline \textit{\small Advantages:} \newline \tabitem Low semantic extraction complexity and latency}
\newline \textit{\small Limitations:} \newline \tabitem Consideration of semantic importance solely from the temporal domain \newline \tabitem Lack of semantic recovery process {\color{Reds}\newline \tabitem Potential semantic information loss} {\quad} \newline \textit{\small Representative Refs:} \cite{zhao2023age,pappas2021goal,9921185,holm2021freshness,nikkhah2023age} & {\quad} \newline \vspace{ -0.6cm}\newline \textit{\small Achieved Capabilities} \newline  \tabitem Semantic Extraction \& Generation {\quad} 
{\color{Reds}\newline \textit{\small Advantages:} \newline \tabitem High quality of generated data \newline \tabitem Low compression rate via cross-modality \newline \tabitem Decoupled paradigm for flexible deployment}
\newline \textit{\small Limitations:} \newline \tabitem No consideration of redundancy in time domain \newline \tabitem Lack of awareness of decision-making {\quad} \newline \textit{\small Representative Refs:} \cite{nam2023language, grassucci2023generative, feng2023scalable, raha2023generative,grassucci2023diffusion}\\ \hline
\multicolumn{3}{|c|}{\textbf{Proposed Agent-Driven Generative SemCom}} \\ \hline
\multicolumn{3}{|p{16.8cm}|}{{\quad} \newline \vspace{ -0.6cm} \newline \textit{\small Achieved Capabilities:}  \qquad \tabitem Joint Semantic Extraction and Sampling \& Inference and Interpolation \vspace{0.05cm} \newline \textit{\small{Advantages:}} \quad  \quad  \tabitem Efficient semantic compression applied  to both each individual data unit and the temporal correlation\newline  \textit{\color{white}\small Superiority:}\qquad \;  \tabitem Adaptive semantic extraction and sampling in response to changes in the channel gain and source content \newline  \textit{\color{white}\small 
 Superiority:}\qquad \;  \tabitem Seamless integration with existing mature digital transmission  for achieving remote communication \newline  \textit{\color{Reds}\small{Limitations:}}\qquad \;  {\color{Reds}\tabitem Only applicable to scenarios where the transmitted content is regular and predictable} \newline  \textit{\color{white}\small Limitations:}\qquad \;   {\color{Reds}\tabitem Periodic fine-tuning is required to adapt to changes in the communication context}} 

\\ \hline
\end{tabular}
 
   \label{Comparison}

\vspace{-0.5cm}
\end{table*}

\section{Related works}
\label{sec:related work}
In this section, we first introduce the development of two mainstream research directions in SemCom, namely semantic-extraction based and semantic-sampling based SemCom. We then present the vision of a new SemCom paradigm enabled by GAI, demonstrating how GAI techniques can effectively combine semantic extraction and sampling. This sets the foundation for the A-GSC framework proposed in this work.

\subsection{Semantic {Extraction} based SemCom}
In the research of SemCom that focuses on semantic extraction, DL stands out as the predominant technology, effectively applied to various types of modal data~\cite{qin2021semantic}. Beyond the widely proven efficient semantic compression capability, recent advancements have expanded its applicability to more intricate and practical scenarios. For image transmission, the authors of~\cite{zhang2023predictive, yang2024swinjscc} introduce an adaptive deep coding framework within the context of variable code length enabled DL-based joint source-channel coding (DeepJSCC). This framework addresses specified target transmission quality requirements by incorporating an Oracle Network-based PSNR (peak signal-to-noise ratio) module. Furthermore, the authors in~\cite{wang2022wireless} leverage a learned nonlinear transform function to establish a temporally adaptive entropy model for customizing DeepJSCC for video. Additionally, addressing specific cases, the authors of~\cite{zhang2022deep} and~\cite{zhang2023deepma} propose enhanced DeepJSCC frameworks for SemCom systems with task-unaware transmitters and those with multiple access, respectively.

Despite these advancements, all DeepJSCC frameworks are trained in an end-to-end manner. The non-explainability hinders the semantic information to be interpreted or manipulated as intended, e.g., the semantic sampling as discussed in Section~\ref{sec:2-A}. Moreover, the most crucial point is that DeepJSCC  only shows significant performance advantages in analog  transmissions~\cite{xie2021deep}, and its effectiveness heavily depends on the match between the actual  and  training channel environment. These limitations restrict its effectiveness to proximity single-hop links
without relay forwarding, making it difficult to apply in the remote monitoring scenarios that
may involve the core network and multi-hop forwarding. To tackle these challenges, some researchers employ knowledge graphs (KG) as semantic information containers, performing semantic encoding and decoding based on KG generation, KG embedding, and KG embedding reasoning~\cite{yang2023task, farshbafan2023curriculum}. However, the widespread adoption and advancement of this approach are hindered by its high computational complexity. 
\subsection{Semantic Sampling based SemCom}
\label{sec:2-A}
Replacing the metric of delay with Age of Information (AoI) in network optimization signifies a pivotal initial step in the transition from conventional communication to semantic communication (SemCom). This transition is crucial for delineating the semantic significance of diverse data in the temporal domain~\cite{9919752}. Addressing the stochastic nature of environments, various AoI-based sampling and scheduling algorithms have been explored across different scenarios. These include optimization for time-average age-based sampling~\cite{9380899}, peak-age-based sensing~\cite{10354523}, and age-threshold-based access~\cite{zhao2023age}. Despite these efforts, AoI alone may not sufficiently address task-specific requirements~\cite{ayan2019age}.

The aforementioned algorithms, being data-content-agnostic, run the risk of redundantly transmitting irrelevant data in certain scenarios~\cite{pappas2021goal}. In response to this challenge, some studies have integrated data content and task requirements into a novel metric for AoI, termed Value of Information (VoI). VoI is conceptualized as a non-linear function of AoI, taking different forms for distinct communication systems. For systems ensuring real-time updates at the receiver regarding the source's status, VoI corresponds to the age of incorrect information~\cite{9921185}. In a pull-based system reliant on query-driven communication, VoI is specified as the age of information at the query~\cite{holm2021freshness}. Additionally, for systems incorporating wireless power transfer, where the receiver must act upon the received status, VoI is adapted to represent the age of actuation~\cite{nikkhah2023age}.

Nonetheless, existing research has overlooked the semantic information embedded in source data. Particularly in monitoring scenarios, visual data often contains redundant information. Consequently, there is a compelling need for an integrated approach that combines semantic extraction and sampling.

\subsection{Prompt Extraction based SemCom}
Recently, the emergence of GAI presents a promising opportunity to revolutionize the SemCom framework. The focus has shifted from enhancing the efficiency of semantic encoding to the extraction of precise prompts. In this paradigm, the shared knowledge base (KB) of both communication participants can be more effectively leveraged~\cite{ren2023knowledge}. Unlike the DeepJSCC paradigm, where KB only contributes to the training process of the semantic encoder and decoder, the generative SemCom paradigm allows background information to play a crucial role in semantic inference during each communication instance. This capability helps reduce the volume of information that needs to be transmitted. 

Pioneering efforts in Generative SemCom have explored applications in language-oriented communication~\cite{nam2023language}, image transmission~\cite{grassucci2023generative, feng2023scalable, raha2023generative}, and audio transmission~\cite{grassucci2023diffusion}. These endeavors have demonstrated high-quality reconstructed data and the fidelity of semantic information. However, the full potential of GAI for SemCom remains largely unexplored. The explainability allows the independent design and fine-tuning of transceivers. This, in turn, facilitates the seamless integration of new functional modules into the modular system. Moreover, 
it removes the need for channel modeling during model training. It can further leverage existing source coding technologies and integrate with current digital transmission networks to address complex and diverse network environments.

\section{System Model and Overview}
\label{sec:systemovervive}
Without loss of generality,  our focus centers on  a common remote  surveillance scenario in intelligent
transportation systems\footnote{An example of {OneMotoring by Land Transport Authority} can be found at \url{https://onemotoring.lta.gov.sg/content/onemotoring/home/driving/traffic_information/traffic-smart.html}.}. The source in the considered scenario is an embedded vision sensor characterized by limited memory and computational capabilities, which is responsible for monitoring a certain scene,  and updating the destination on the scene changes in a timely manner. The destination is a remote server with \textit{ predictable characteristics}, which is tasked with reconstructing the remote real-time scene, with the objective of presenting users with the illusion of real-time transmission.
We assume that SemCom technology is deployed within the system. The system model is specifically described from the following three aspects.

\begin{figure*}[t]
 \centering
\includegraphics[width=1\linewidth]{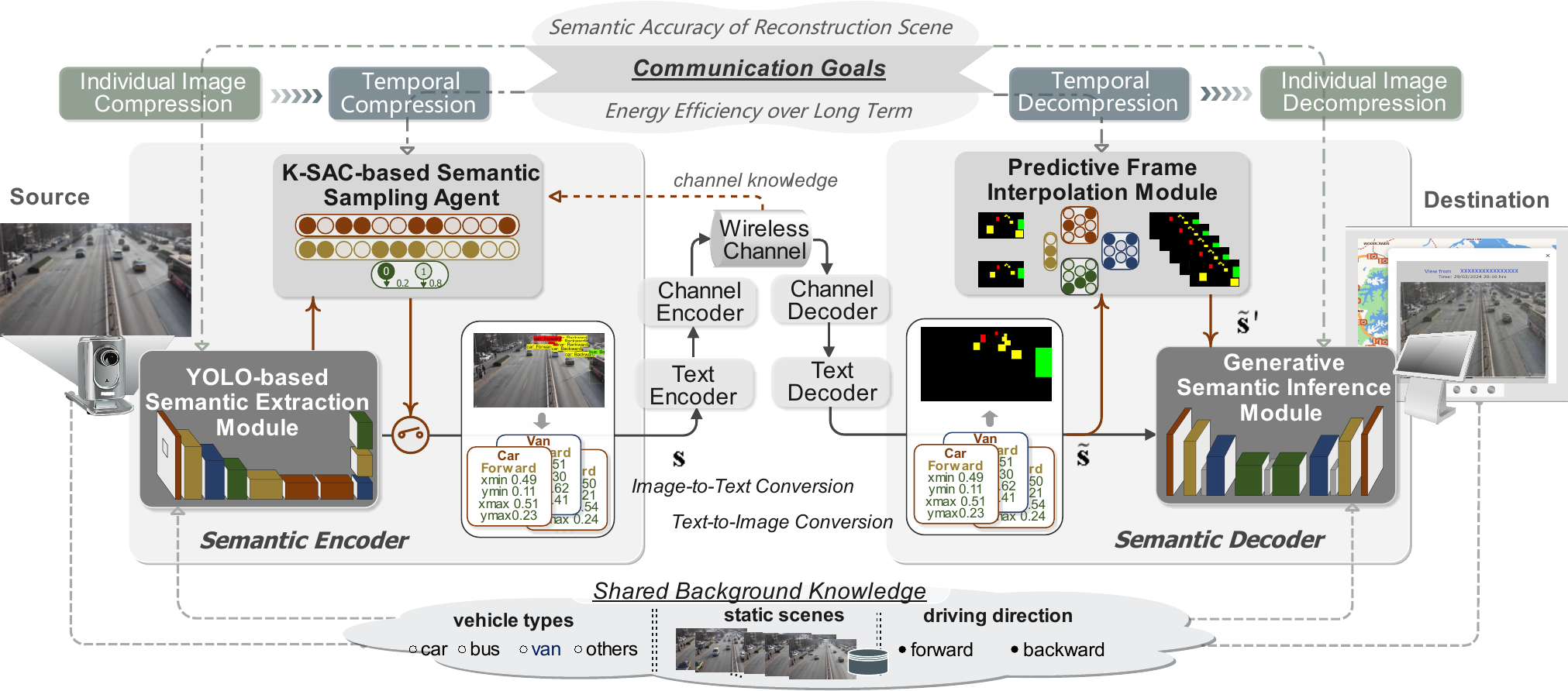}\\
 \caption{An agent-driven generative semantic communication framework
with cross-modality and prediction. 
 }
\label{fig:Framework}
\vspace{-0.5cm}
\end{figure*}
\subsection{Shared Knowledge Base Model for SemCom}
\label{sec:BK model}

The shared KB is introduced as a pivotal component in SemCom systems, providing prior side information to enhance semantic encoding~\cite{ren2023knowledge}. Unlike traditional DL-based SemCom approaches that non-selectively incorporate all KB into the empirical dataset for end-to-end training, the proposed generative SemCom framework strategically analyzes KB to unlock its full potential in boosting SemCom performance. 

In the specific context considered, it can be universally agreed that the destination is solely concerned with the vehicular traffic on the road. In this sense, by utilizing shared historical surveillance data, both sources and the destination can simplify and standardize mutually agreed-upon the essential semantic representations, including vehicle location, type, and driving direction. Meanwhile, since the static background information of the monitoring scene, e.g., the view layout, as well as dynamic changes involving irrelevant pedestrians and non-motorized vehicles along the roadside, are of no interest to the destination, they can be disregarded by the source before transmission and then reconstructed at the destination based on the historically shared knowledge base, despite potential slight differences. 
Based on the guidance provided by the analysis of KB for specific tasks as described above, we  embark on the customized design of the SemCom system.

\subsection{ Semantic Processing Model for Source and Destination}

Based on the task-specific KB analysis, to compress the data for transmission,  a \textit{semantic extraction module} is utilized to extract information on vehicular traffic in the monitoring scene. The destination requires  a \textit{semantic inference module}, which reconstructs the remote monitoring scene solely from the updated traffic information. Naturally, such scene reconstruction depends on shared KB. In other words, the shared KB within this framework acts as a bridge between the source and destination, decoupling the design of the encoder and decoder.

In addition, considering the communication objectives of the surveillance task, the destination is more interested in changes in the remote scene, specifically  in traffic flow rather than in background information. In this sense, performing appropriate sampling can further reduce the amount of data to be transmitted from the temporal dimension.
Thus, we incorporate a \textit{semantic sampling agent} at the source to perform a binary 0-1 action—deciding whether to transmit the current semantic information or not, with the consideration of  the semantic changes in the monitoring scene and other factors. 
 Additionally, we integrate a \textit{predictive frame
interpolation module} at the destination, enabling the  illusion of real-time transmission to users, while also potentially reducing the number of required sampling instances. 
The specific designs of the involved modules related to the semantic extraction and semantic sampling  are detailed in Section~\ref{sec:compression} and Section~\ref{sec:sampling}.


\vspace{-0.2cm}
\subsection{Wireless Transmission Model}
\label{sec:channel}
In this work,  we believe that wireless transmission is a bottleneck in the communication process. As such, we mainly analyze and evaluate the wireless transmission performance. 
Considering the combined effects of multi-path and shadowing on the practical transmission, we adopt the $\mathcal{F}$ composite
fading model to characterize the stochastic wireless channel~\cite{8638956}. We denote
the instantaneous channel gain by  $\tilde g$. The probability density function of  $\tilde g$ is given by~\cite{8638956}
\begin{equation}
    f\left( {\tilde g} \right) = \frac{{{m^m}{{\left( {{m_s} - 1} \right)}^{{m_s}}}{{\bar g}^{{m_s}}}{{\tilde g}^{m - 1}}}}{{B\left( {m,{m_s}} \right){{\left[ {m\tilde g + \left( {{m_s} - 1} \right)\bar g} \right]}^{m + {m_s}}}}},\label{eq:pdf}
\end{equation}
where $m$ and $m_s$ represent the number of clusters of multipath, shadowing shape, respectively, and ${\bar g}$ is the corresponding average channel gain, i.e., $\bar g = \mathbb{E}\left[ {\tilde g} \right]$. Moreover, $B\left( { \cdot , \cdot } \right)$ denotes the beta function~\cite{8638956}. 
Since we focus on the design of the semantic encoder and decoder, without loss of generality, we assume perfect
capacity achieving coding in this work. Moreover, in order to cope with stochastic fading, it is assumed that the transmitter of the embedded vision sensor adopts a power control technique. We denote the decoding threshold for the signal-to-noise ratio by $\Theta $.
Thus, the achievable transmission rate is expressed by
\begin{equation}
 {R} = W\log \left( {1 + \Theta } \right),   
\end{equation}
 where $W$ is the allocated bandwidth. The instantaneous transmit power is then given by
$\tilde p = {{\Theta {\sigma ^2}} \mathord{\left/
 {\vphantom {{\Theta {\sigma ^2}} {\tilde g}}} \right.
 \kern-\nulldelimiterspace} {\tilde g}}$, where ${\sigma ^2}$ represents noise power.

\begin{figure}
    \centering
\includegraphics[width=0.8\linewidth]{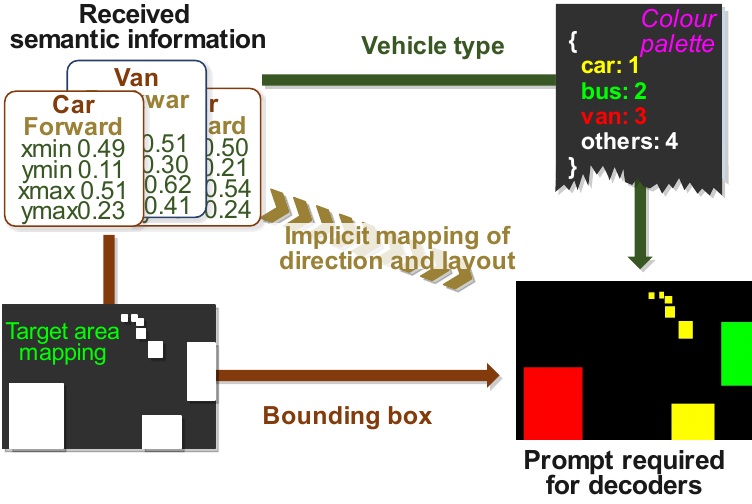}
    \caption{Transformation from semantic information to prompt.}
    \label{fig:transform}
    \vspace{-0.5cm}
\end{figure}

\section{Key Component Design: Semantic Extraction} 
\label{sec:compression}
In this section, we focus on semantic extraction, a key aspect of SemCom. Within the proposed framework, semantic extraction and inference are achieved through a YOLO-based semantic extraction module and generative semantic inference modules, respectively.

\subsection{Semantic Extraction Module}\label{sec: extraction}
Based on the strategic analysis on the shared  KB as stated in Section~\ref{sec:BK model}, we can see that the destination is mainly concerned with the information about the distribution of the traffic flow. To track the vehicles' information in real time, the semantic extraction can be performed based on a mature real-time object detection algorithm, such as the low-complexity YOLOv5~\cite{liu2023yolov5sbc} which can be deployed in multiple embedded and mobile devices. Furthermore, we characterize the semantic information about the position for vehicle $v$ as the bounding box, which is denoted by a 4-tuple {\color{Reds}${\mathbf{b}}_v = \left( {{b_{1}^v},{b_{2}^v},{b_{3}^v},{b_{4}^v}} \right) \in {\left[ {0,1} \right]^4}$. Therein, $\left( {b_1^v,b_2^v} \right)$ represents the coordinate of the top-left corner, and $\left( {b_3^v,b_4^v} \right)$ represents the coordinate of the bottom-right corner.}
Given that for a two-way road, the driving directions of the vehicles are tightly coupled with their locations, we do not consider the semantic information of the vehicle direction, separately. To avoid the ambiguous semantic information of the bounding box arising from the perspective effect of the position of the vehicle and the size of the vehicle type itself, we use a scalar to represent the type of vehicle, denoted by $o_v \in \mathcal{O} = \left\{ {{1}, \ldots ,{k}, \ldots ,{K}} \right\}$, where $K$ represents the total number of the vehicle types. In summary, a piece of semantic information for a vehicle $v$ in the monitoring scene can be represented by a vector ${{\mathbf{s}}_v} = \left[ { o_v,{{\mathbf{b}}_v}} \right]$, the data size of which after text data encoding is denoted by $\ell$\footnote{We have total vehicle types $K=4$, and use two bits to characterize ${o_v}$, and use five bits to characterize $x_{\min }^v$, $y_{\min }^v$, $x_{\max }^v$, and $y_{\max }^v$, respectively, with a resolution of 0.03125. Thus, we have  $\ell = \left( {2 + 4 \times 5} \right)$ bit.}. Assuming a total of $M$ vehicles in the scene, the total semantic information contained in the scene can be represented as ${\mathbf{s}} = \left[ {{{\mathbf{s}}_1}, \ldots ,{{\mathbf{s}}_v}, \ldots ,{{\mathbf{s}}_M}} \right]$. Before transmission, the semantic information is packed into a packet with the data size of $L=M\ell$. If there are no vehicles in the monitored scene, an empty semantic message will be sent to the receiving side. Transforming the originally captured scene images into textual layout information  can significantly enhance the compression ratio at the source side. For a relevant quantitative comparison with other representative SemCom paradigms, please refer to our prior research~\cite{wang2024harnessing}.

\begin{figure*}[t]
 \centering
\includegraphics[width=0.9\linewidth]{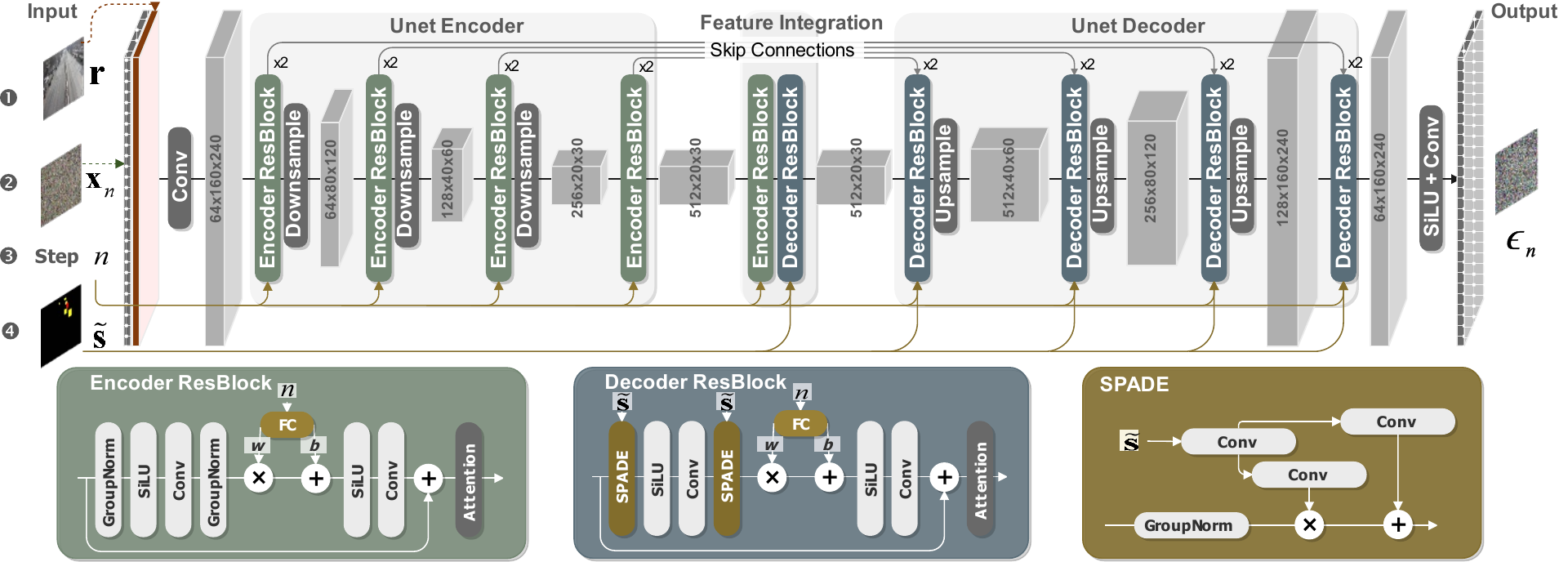}\\
 \caption{U-Net neural network  for noise prediction~\cite{park2019semantic,wang2022semantic}. 
 }
\label{fig:Unet}
\vspace{-0.5cm}
\end{figure*}
\subsection{Semantic Inference Module}
Given the requirement for photorealistic scene reconstruction, we utilize the diffusion model as the foundational element of our semantic inference module design. However, multimodal fusion of the textual prompts and image features during the diffusion process presents a significant challenge. In existing layout diffusion models, issues such as object omission can occur when the number of objects is excessively large or their size is too small~\cite{zheng2023layoutdiffusion}. This mainly arises from the absence of structured spatial information in the textual layout.

To address this, we do not directly use the received textual semantic information as the prompt for scene reconstruction. Instead, we first visualize the textual information based on knowledge base as shown in Fig.~\ref{fig:transform}, thereby alleviating the burden of the cross-modal fusion in the diffusion model. Through target area mapping, the position and size of the vehicles can be easily fused with the positional information of each feature map during the diffusion process. \textit{This not only addresses the challenge of multimodal fusion between text and images in existing layout diffusion research but also removes the limitation of the existing layout diffusion models  on the maximum number of objects.} Additionally, to differentiate the specific categories of objects, we paint the areas corresponding to different vehicles with resolution values.  Meanwhile, for synthesizing a seamless and realistic scene, we also use the static scene information of the remote scene as a prompt for the diffusion model, which is denoted by ${\mathbf{r}}$. It should be noted that the static scene information does not need to be updated in real-time to the destination, unless there are changes in the weather or times of day. This is because  the dynamics in the background of exclusion zones are not of concern.

 The diffusion model itself is a latent variable model, which consists of a forward and a reverse process. 
 We denote the noisy data at denoising step $n$ by ${{\mathbf{x}}_n}$.
 The \textit{forward} process is defined as a discrete Markov chain of length $N$: $q\left( {{{\mathbf{x}}_{1:N}}\left| {{{\mathbf{x}}_0}} \right.} \right) = \prod\nolimits_{n = 1}^N {q\left( {{{\mathbf{x}}_n}\left| {{{\mathbf{x}}_{n - 1}}} \right.} \right)}$. In each step $n \in \left[ {1,N} \right]$ in the forward process,  the diffusion model adds noise ${\epsilon _n}$ sampled from the standard Gaussian distribution $\mathcal{N}\left( {0,{\mathbf{I}}} \right)$ to data ${{{\mathbf{x}}_{n-1}}}$ and obtains disturbed data ${{{\mathbf{x}}_{n}}}$ from $q\left( {{{\mathbf{x}}_n}\left| {{{\mathbf{x}}_{n - 1}}} \right.} \right) = \mathcal{N}\left( {{{\mathbf{x}}_n};\sqrt {1 - {\beta _n}} {{\mathbf{x}}_{n - 1}},{\beta _n}{\mathbf{I}}} \right)$, where ${{\beta _n}}, n \in \left[ {1,N} \right]$ characterizes the scale of added noise at each step $n$. Notably, during the training process, we can sample ${{{\mathbf{x}}_n}}$ at any step $n$ in closed form via $q\left( {{{\mathbf{x}}_n}\left| {{{\mathbf{x}}_0}} \right.} \right) = \mathcal{N}\left( {{{\mathbf{x}}_n};\sqrt {{{\bar \alpha }_n}} {{\mathbf{x}}_0},\left( {1 - {{\bar \alpha }_n}} \right){\mathbf{I}}} \right)$, with ${\bar \alpha _n}: = \prod\nolimits_{s = 1}^n {\left( {1 - {\beta _s}} \right)}$.
Moreover, given the prompts ${{\mathbf{\tilde s}}}$  and ${\mathbf{r}}$ feed into the diffusion model, the reverse process can also be defined as a Markov chain: ${p_\theta }\left( {{{\mathbf{x}}_{0:N}}\left| {{\mathbf{\tilde s}},{\mathbf{r}}} \right.} \right) = p\left( {{{\mathbf{x}}_N}} \right)\prod\nolimits_{n = 1}^N {{p_\theta }\left( {{{\mathbf{x}}_{n - 1}}\left| {{{\mathbf{x}}_n},{\mathbf{\tilde s}},{\mathbf{r}}} \right.} \right)}$, where 
$ {{p_\theta }\left( {{{\mathbf{x}}_{n - 1}}\left| {{{\mathbf{x}}_n},{\mathbf{\tilde s}},{\mathbf{r}}} \right.} \right)}$ can be parameterized  as $\mathcal{N}\left( {{{\mathbf{x}}_n};{\mu _\theta }\left( {{{\mathbf{x}}_n},n} \right),{\sigma _n}} \right)$, where ${{p_\theta }\left( {{{\mathbf{x}}_{n - 1}}\left| {{{\mathbf{x}}_n},{\mathbf{\tilde s}},{\mathbf{r}}} \right.} \right)}$ is parameterized as $\mathcal{N}\left( {{{\mathbf{x}}_n};{\mu _\theta }\left( {{{\mathbf{x}}_n},{\mathbf{\tilde s}},{\mathbf{r}},n,} \right),{\sigma _n}} \right)$. By modeling and comparing the posterior distribution of the forward process, the mean term can be further rearranged as ${\mu _\theta }\left( {{{\mathbf{x}}_n},{\mathbf{\tilde s}},{\mathbf{r}},n,} \right) = \frac{1}{{\sqrt {1 - {\beta _n}} }}\left( {{{\mathbf{x}}_n} - \frac{{{\beta _n}}}{{\sqrt {1 - {{\bar \alpha }_n}} }}{\epsilon _\theta }\left( {{{\mathbf{x}}_n},{\mathbf{\tilde s}},{\mathbf{r}},n} \right)} \right)$, and the variance term can be approximated as ${\sigma _n} = {\beta _n}{\mathbf{I}}$ without sacrificing performance. Furthermore, ${\epsilon _\theta }$ is parameterized by a U-Net network (as shown in Fig.~\ref{fig:Unet}) of which inputs are ${{{\mathbf{x}}_n}}$, ${{\mathbf{\tilde s}}}$, ${\mathbf{r}}$, and $n$. In light of above, the simplified loss function guiding the training can be expressed as
\begin{equation}
    {\mathcal{L}_\theta } = {\mathbb{E}_{n,{{\mathbf{x}}_n},}}\left[ {{{\left\| \epsilon-{ \epsilon {_\theta }\left( {\sqrt {{{\bar \alpha }_n}} {{\mathbf{x}}_n} + \sqrt {1 - {{\bar \alpha }_n}}\epsilon ,{\mathbf{\tilde s}},{\mathbf{r}},n]} \right)} \right\|}_2}} \right].
\end{equation}
Then, during the inference process, samples are generated by iterating through the reverse process ${p_\theta }\left( {{{\mathbf{x}}_{0:N}}\left| {{\mathbf{\tilde s}},{\mathbf{r}}} \right.} \right)$ from $n=N$ to $n=0$. Specifically, ${{\mathbf{x}}_N} \sim \mathcal{N}\left( {0,{\mathbf{I}}} \right)$, and ${{{\mathbf{x}}_n}}$ in each step is predicted as 
\begin{equation}
    {{\mathbf{x}}_{n - 1}} = \frac{1}{{\sqrt {1 - {\beta _n}} }}\left( {{{\mathbf{x}}_n} - \frac{{{\beta _n}}}{{\sqrt {1 - {{\bar \alpha }_n}} }}{\epsilon _\theta }\left( {{{\mathbf{x}}_n},{\mathbf{\tilde s}},{\mathbf{r}},n} \right)} \right) + {\sigma _n}{\mathbf{z}},
\end{equation}
where ${\mathbf{z}} \sim \mathcal{N}\left( {0,{\mathbf{I}}} \right)$\footnote{The detailed derivation and proof can be found in~\cite{ho2020denoising}.}.
\section{Key Component Design: Semantic Sampling} 
\label{sec:sampling}
This section shifts focus to semantic sampling, aiming to streamline transmission in the \textit{temporal} domain, particularly like \textit{real-time} operations such as surveillance. To achieve this, we place a RL-based \textit{semantic  sampling agent} at the source side to solve the sequential decision problem.  To ensure that the sampling behavior is unnoticeable to the  destination, a \textit{predictive frame interpolation module} must be pre-installed on the destination side. When the agent and the  module work in harmony, the SemCom system's performance evaluation will align with the underlying SemCom framework given in Section~IV.
We next begin with the design of the semantic sampling agent, followed by a detailed illustration of the involved predictive frame interpolation module.

\vspace{-0.2cm}

\subsection{K-SAC Semantic Sampling Agent}\label{sec:agent}
The semantic sampling agent is {\color{Reds}responsible for adaptively performing semantic sampling based on varying channel conditions and source data content}, with the objective of minimizing the amount of transmitted data while ensuring optimal performance in SemCom. 
Although the RL paradigm can address the challenge of the intricate correlation between the \textit{\color{Reds}state} and the \textit{\color{Reds}reward}, an issue remains. Prior simulations reveal that prediction deviation and sampling interval are not strictly linearly related due to the influence of object size and object moving speed. To mitigate the ambiguity that the nonlinearity creates for the agent's learning, we adopt a \textit{stochastic} learning policy named soft actor-critic (SAC), making it robust to such occasional nonlinearities.
Additionally, to avoid the impact of convergence latency on communication performance and the communication overhead resulting from immediate reward feedback from the destination to the source, we integrate knowledge into the SAC algorithm~\cite{she2021tutorial}. By modeling wireless transmission using the acknowledged empirical channel model, $\mathcal{F}$ composite fading model, virtual experiences can be generated to facilitate offline training. Thus, we refer to our algorithm as K-SAC, and its MDP model is detailed as below. 

\begin{algorithm*}
\footnotesize
Initialize reply memory ${\mathcal{D}}$; Initialize  weights of actor network and critic networks\\
\For{episode j = 1 to MAX\_EPISODE}{
Create a buffer  ${{\mathbf{f}}_j}$  for recording the visual layout  sequence at the destination \\
Predict the next $P$ visual layouts and create a list of variable length denoted by ${\mathbf{\tilde s}}' = \left[ {{\mathbf{\tilde s}}{'_1},{\mathbf{\tilde s}}{'_2}, \ldots ,{\mathbf{\tilde s}}{'_P}} \right]$ \\
Calculate semantic change degree $\chi _t$ based on ${{\mathbf{s}}_{t}}$ and $ {{\mathbf{s}}_{\hat t}}$ \\
\textit{Randomly} initialize the initial state ${{\mathbf{S}}_1}$ \\
\For{$t = 1$ to $T$ (MAX\_STEP)}{
\eIf{${\mathbf{\tilde s}}'$ is non-empty}{Pop up the predicted visual layout ${{\mathbf{\tilde s}}{'_t}}$ from the head of the list ${\mathbf{\tilde s}}'$}{Perform next loop visual layout prediction based on  previously popped up visual layouts ${\mathbf{\tilde s}}{'_{t - 1}}$ and ${\mathbf{\tilde s}}{'_{t - 2}}$  }
Generate an action $a_t$ based on the probability distribution output by the current policy $\pi_t$  \\
\eIf{Perform semantic sampling (${a_t} = 1$)}{
Transform semantic information ${{\mathbf{s}}_t}$ into visual layout ${{{\mathbf{\tilde s}}}_t}$ and store the real visual layout ${{{\mathbf{\tilde s}}}_t}$ in to ${{\mathbf{f}}_j}$ \\
Perform the prediction for the next $P$ visual layouts based on ${{{\mathbf{\tilde s}}}_t}$ and last sampled ${{{\mathbf{\tilde s}}}_{\hat t}}$ and update list ${\mathbf{\tilde s}}'$\\ Estimate the energy consumption $E_t$ according to \eqref{sm}}{Store the predicted visual layout ${\mathbf{\tilde s}}{'_t}$ in to ${{\mathbf{f}}_j}$ \\Compare the deviation $D_t$ between ${\mathbf{\tilde s}}{_t}$ and ${\mathbf{\tilde s}}{'_t}$ according to \eqref{eq:Dt}\\
 } 
Calculate reward $r_t$ according to \eqref{eq:G} and ${{\mathbf S}_{t + 1}} \leftarrow {{\mathbf S}_t}$; Record transition ${e_t} = \left[ {{\mathbf{S}_t},{a_t},{\mathbf{S}_{t + 1}},r} \right]$ in ${\mathcal D}$\\
Sample a batch of $\mathcal{B}$ to perform a gradient descent according to \eqref{eq:L1}, \eqref{eq:L2}, and \eqref{eq:L3}\\
Update the parameters following: $\theta  \leftarrow \theta  - {\lambda _\theta }\nabla {L_Q}\left( \theta  \right)$; $\phi  \leftarrow \phi  - {\lambda _\phi }\nabla {L_\pi }\left( \phi  \right)$; $\vartheta  \leftarrow \vartheta  - {\lambda _\vartheta }\nabla L\left( \vartheta  \right)$
}
}

\caption{K-SAC-based Agent-driven Semantic Sampling Algorithm}
\label{alg:Agent}
\vspace{0cm}
\end{algorithm*}

\subsubsection{MDP model} 
\label{sec:MDP}
Specifically, the time step in RL framework is specified as a sensing time interval (STI). The duration of STI is determined by the supportable frame rates for vision sensors and the inference time of the diffusion-based generation module, and each STI is indexed by $t$. As shown in Fig.~\ref{fig:KSAC}, the entire SemCom process can be viewed as the environment. {\color{Reds}The semantic sampling agent corresponds to the agent in an MDP,} which interacts with the environment by performing sampling actions. The goal of the agent is to achieve the finest cumulative reward $\hat r$ within the observation window (denoted by $W$) with a duration of $T$. Denote the immediate reward by $r_t$. We  define $\hat r = \mathbb{E}\left[ {\sum\nolimits_{t = 0}^T {{\gamma ^t} \cdot {r_t}} } \right]$, where $\gamma$ is the discount factor.
Since we aim to achieve consistently high scene reconstruction accuracy,  $\gamma$ is set to 1.
The interactions between the agent and the knowledge-based virtual environment are presented as below. 

\textbf{State:} The state observed by the agent at time step $t$ is denoted as ${\bf S}_t$, encompassing four critical factors: specifically, ${{\mathbf{S}}_t} = \left[ { {L_t}, {\bm\chi _t}, {{\bar g}_t}} \right]$.
Herein, 
$L_t$ characterizes the current data size of the semantic information packet as defined in Section~\ref{sec: extraction}. 
Furthermore, ${\bm\chi _t} = \left[ {{\chi _t},{\chi _{t - 1}}, \ldots ,{\chi _{t - W}}} \right]$ conveys historical insights into the semantic change degree in the past $W$ STIs, facilitating the agent to capture the urgency associated with the sampling process. In the considered task, 
We denote the timestamp corresponding to the last sampled scene  by $\hat t$. Then, the semantic change degree is measured based on the cumulative deviation of bounding boxes for each vehicle in the current scene compared to the last sampled scene, while irrelevant information, such as time of day and weather, can be ignored. 
Specifically, ${{\chi _t}}$ can be calculated according to semantic information ${{\mathbf{s}}_t}$ and ${{\mathbf{s}}_{\hat t}}$ as
\begin{equation}
   {\chi _t} = \sum\limits_{v = 1}^M {\frac{{A_v^t + A_v^{\hat t} - 2{I_v}}}{{2(A_v^t + A_v^{\hat t} - {I_v})}}}  , \label{change}  
\end{equation}
where {\color{Reds}${I_v} = |\min \{ b_{3}^{v,t},b_{3}^{v,\hat t}\}  - \max \{ b_{1}^{v,t},b_{1}^{v,\hat t}\} | \cdot |\min \{ b_{4}^{v,t},b_{4}^{v,\hat t}\}  - \max \{ |b_{2}^{v,t},b_{2}^{v,\hat t}\} |$ and $A_v^t = |b_{2}^{v,t} - b_{2}^{v,t}| \cdot |b_{4}^{v,t} - b_{4}^{v,t}|$.}
At last, ${{\bar g}_t}$ denotes the average channel gain in time step $t$, which can provide a basis for the agent to assess the energy consumption attributed to the sampling operation. 
{\color{Reds}The agent perceives the state of environment and uses the information to make the decision as presented as below. Afterward, state ${\bf S}_t$ transitions to the next state, ${\bf S}_{t+1} = \left[ { {L_{t+1}}, {\bm\chi _{t+1}}, {{\bar g}_{t+1}}} \right]$, based on the executed action. Specifically, the agent's actions primarily affect the state element ${\chi _{t + 1}}$ in ${{\bf{\chi }}_{t + 1}}$. Both ${L_{t+1}}$ and ${{\bar g}_{t+1}}$ are equivalent to the parameters of the MDP and are directly determined by the size of the semantic information of the source data and the channel conditions  at STI $t +1$, independent of the action outcome.}
It is essential to emphasize that, in this work, we assume that the source is stationary. In cases involving moving sensors, the state should encompass the historical channel variations to capture the influence of user mobility on transmission energy consumption.

\textbf{Action:} The action  performed by the semantic sampling agent at the beginning of time step $t$ can be expressed by $a_t \in \mathcal{A} = \{0, 1\}$. If $a_t = 1$, semantic sampling takes place within STI $t$. {\color{Reds}Meanwhile, the timestamp corresponding to the last sampled scene is updated as   
$\hat t \leftarrow t$, which  intervenes in the state transitions as stated above.}  Otherwise, semantic sampling is not carried out.

\textbf{Reward:}
The reward acquired by the agent at the end of STI $t$ is denoted by $r_t$. In the pursuit of highly efficient remote  surveillance, the immediate reward is considered with regard to two key factors. The first factor pertains to the precision of the reconstructed scene measured inversely through the prediction deviation  $D_t$ between the real-time visual layout ${{\mathbf{\tilde s}}_t}$ and the predicted visual layout ${{\mathbf{\tilde s}'}_t}$. Note that the prediction method is detailed in Section~\ref{SEC:PREDICTION}. 
The second factor relates to the energy consumption $E_t$ incurred during the transmission of the sampled semantic map. With  both factors in mind, two sub-reward functions are designed: one for scenarios where semantic sampling is performed, denoted by ${r_t^{\text 1}}$
, and the other for scenarios without semantic sampling, denoted by ${r_t^{0}}$.
Then, the immediate reward function can be expressed as follows:
\begin{equation}
    {r_t} =  {r_t^{\text 1}}{{\mathbf{1}}_{{a_t} = 1}} + {r_t^{\text 0}}\left( {1 - {{\mathbf{1}}_{{a_t} = 1}}} \right). \label{eq:G}
\end{equation}

{\color{Reds}It is worth noting that, in the conventional RL paradigm, the reward is obtained through the agent's interaction with the actual environment. However, in K-SAC, to reduce communication overhead, we model the interaction between the agent and the environment based on existing channel modeling and mathematical analysis, generating virtual experiences (${\bf S}_t, a_t, r_t, {\bf S}_{t+1}$) that can be used for agent training.}
The specific forms and design of ${r_t^{\text 1}}$ and ${r_t^{\text 0}}$ are as follows:

\begin{figure}[t]
 \centering
\includegraphics[width=1\linewidth]{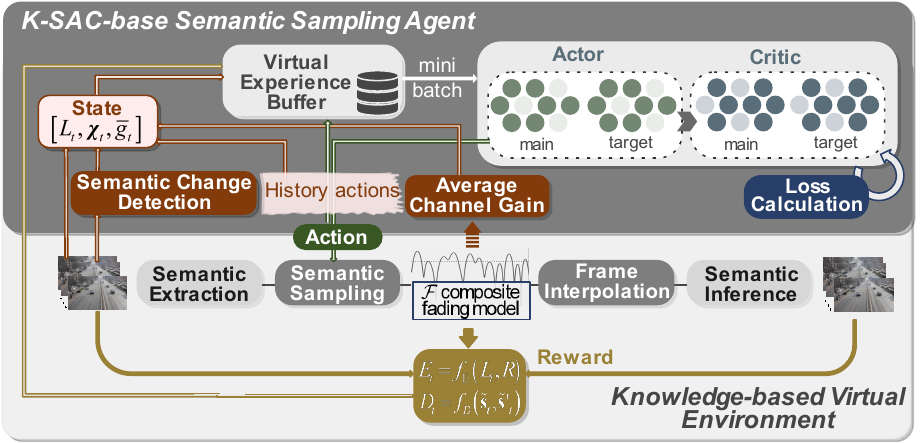}\\
 \caption{K-SAC based semantic sampling agent design.
 }
\label{fig:KSAC}
\vspace{-0.4cm}
\end{figure}

Firstly, to minimize unnecessary sampling, we define the sub-reward $r_t^1$ for the action of sampling as a function inversely proportional to energy consumption.  {\color{Reds}Considering that the energy consumption per transmission typically ranges around 0.015 mJ, we introduce a weight parameter 
$w_1$ to adjust its influence on the sub-reward by one order of magnitude. This ensures that the scaled values are primarily distributed between 0 and 1. Afterwards, given the values of $E_t$ with significantly less variance compared to $D_t$, we employ a logarithmic function to further accentuate the variations between different levels of energy consumption. Meanwhile, this logarithmic transformation helps to prevent potential excessive energy consumption caused by random fading, thus mitigating its impact on the stability of agent learning.}
Additionally, to maintain a balance between the  energy consumption and prediction deviation terms in the final reward function, we incorporate another weighting coefficient, $w_2$, for the logarithmic component. The specific expression for the sub-reward $r_t^1$ is presented as follows,
\begin{equation}
    r_t^1 = {w_2}\log \left( {1 + {w_1}{E_t}} \right),
\end{equation}
where we choose $w_1 = 10$ and $w_2 = -6$. Moreover, $E_t$ is a function of state elements  $L_t$ and ${\bar g}_t$. Instead of interacting with the real environment, we derive the value of $E_t$ based on the well-accepted $\mathcal{F}$ composite fading model
as stated in Section~\ref{sec:channel}. Given that the transmitter adopts the power control technique, the transmission duration for semantic map ${\bf{m}}_t$ can be expressed by ${\delta _t} = {{{L_t}} \mathord{\left/
 {\vphantom {{{L_t}} R}} \right.
 \kern-\nulldelimiterspace} R}$.  Therefore, we have 
$${E_t} = \int_0^{{\delta _t}} {\frac{{\Theta {\sigma ^2}}}{{{{\tilde g}_k}}}dk},$$ where $k$ is the index of the transmission time interval.  Without loss of generality, we assume that the stochastic fading can be treated as independently and identically
distributed (i.i.d.) among transmission time intervals. Then, $E_t$ can be approximated as
\begin{equation}
  \begin{aligned}
     {E_t} &= {\delta _t}\mathbb{E}{_{\tilde g}}\left[ {{ \tilde p}} \right] = \int_0^\infty  {\frac{{{\Theta}{\sigma ^2}}}{{\tilde g}}f\left( {\tilde g} \right)d\tilde g}\\
     &={\delta _t}{{\Theta}{\sigma ^2}}\int_0^\infty  {{{\tilde g}^{ - 1}}f\left( {\tilde g} \right)d\tilde g}  \\
     & = {\delta _t}{{\Theta}{\sigma ^2}}{\mathbb{E}}\left[ {{{\tilde g}^{ - 1}}} \right].\label{sm}
\end{aligned}  
\end{equation}
According to \eqref{eq:pdf}, with the aid of \cite[Eq. (3.194.3)]{zwillinger2007table}, the ${n^{{\rm{th}}}}$ moment of ${\tilde g}$ can be derived as 
\begin{equation}
{\mathbb{E}}\left[ {{{\tilde g}^n}} \right] = \frac{{{{\left( {{m_s} - 1} \right)}^n}{{\bar g}^n}\Gamma \left( {m + n} \right)\Gamma \left( {{m_s} - n} \right)}}{{{m^n}\Gamma \left( m \right)\Gamma \left( {{m_s}} \right)}},  \label{moment}
\end{equation}
 where  $\Gamma \left(  \cdot  \right)$ represents the gamma function. Substituting the case of $n=-1$ in \eqref{moment} into \eqref{sm}, we can obtain the final expression of  ${E_t}$ as
\begin{equation}
    {{E_t}} = \delta_t\frac{{{\Theta}{\sigma ^2}}{m\Gamma \left( {m - 1} \right)\Gamma \left( {{m_s} + 1} \right)}}{{\left( {{m_s} - 1} \right)\bar g\Gamma \left( m \right)\Gamma \left( {{m_s}} \right)}}. \label{ap}
\end{equation}

Secondly, in designing $r_t^0$, to ensure high semantic accuracy for scene reconstruction while promoting the energy-saving awareness of the agent, we introduce a positive constant term, $w_3$, for non-sampling behavior. The value of $w_3$ is determined based on the average energy consumption and the system's emphasis on   $E_t$ and $D_t$. Furthermore,  $D_t$ is calculated similarly to \eqref{change}. We denote the total number of the pixels occupied by each type of vehicles in  ${{\mathbf{\tilde s}}_t}$ and ${{\mathbf{\tilde s}}'}{_t}$  by $n_{v,t}$ and $n'_{v,t}$, respectively. Meanwhile, the  intersection between them is denoted by  ${{{\hat n}_{v,t}}}$. Then, $D_t$ can be expressed by 
\begin{equation}
{D_t} = \sum\limits_{v = 1}^M {\frac{{{n_{v,t}} + n{'_{v,t}} - 2{{\hat n}_{v,t}}}}{{2\left( {{n_{v,t}} + n{'_{v,t}}} \right)}}}. \label{eq:Dt}
\end{equation}
Given the fact that larger deviations result in increasingly severe consequences, we incorporate an exponential function into this sub-reward formulation.
Furthermore, to proactively address the issue of significant cumulative prediction deviation resulting from prolonged reliance on destination-side predictions, 
we define $\hat D_t$ using a penalty function as below, 
\begin{equation}
   {{\hat D}_t} = \left\{ {\begin{array}{*{20}{c}}
  {{D_t}}&{{D_t} \leqslant {D_{{\text{th}}}}} \\ 
  {\min \left\{ {{D_t} + \kappa   ,1} \right\}}&{{D_t} > {D_{{\text{th}}}}} 
\end{array}} \right..
\end{equation}
This is relative to the actual deviation $D_t$, considering that the tolerable prediction deviation is capped by a threshold, $D_{\rm th}$.
Moreover, we  introduce a weight coefficient $w_4$ for ${\hat D}_t$, the value of which is chosen in conjunction with the value of $w_3$. Thus, the expression of $r_t^0$ is presented as  
\begin{equation}
    r_t^0 = {w_3} - \exp \left( {{w_4}{{\hat D}_t} - 1} \right),
\end{equation}
where we set $w_3 = 1$, $w_4 = 2$,  $D_{\rm th} = 0.07$, and $\kappa  = 0.5$. 

The careful selection of aforementioned weight coefficients  collectively guarantees the rationality of rewards and punishments. This mitigates the agent's inclination towards short-term gains due to insufficient sampling. Simultaneously, it also prevents hesitation in performing a necessary sampling given the potential penalties associated with sampling.

{\color{Reds}\textbf{Policy:} The policy (denoted by $\pi$) describes the mapping from a given state to the agent's action selection, which can achieve the maximum long-term rewards.   At the beginning of Section~\ref{sec:agent}, we choose a stochastic policy, named K-SAC, where the agent selects an action based on the action distribution generated by the policy's output,  $\pi \left( {{{\bf{S}}_t}} \right)$. It is important to note again that K-SAC is an enhancement of the SAC algorithm, specifically designed for an offline RL paradigm. By leveraging existing channel modeling research and mathematical analysis to model the wireless transmission process as discussed above, we can generate virtual experiences based on practical history footage clips, such as the relationships $r_t$ and $\bf{S}_{t+1}$, and $\bf{S}_t$ and $a_t$. These virtual experiences are used for offline pre-training and fine-tuning, followed by periodically updating to the sender's semantic sampling module. This not only reduces the communication overhead during training but also avoids impacting actual communication performance during the convergence process. The learning principle is detailed in Section~\ref{sec:K}.

}

\subsubsection{Learning principle} \label{sec:K}
Based on the MDP designed above, we present the learning part of SAC. For illustration, we denote the state-action trajectory under the policy $\pi$ by ${\rho _\pi }$. In the traditional RL algorithm, the fundamental objective of agent is to learn the optimal policy $\pi$ which maximizes the cumulative expected rewards, which is expressed by 
\begin{equation}
    {\pi ^*} = \arg \mathop {\max }\limits_\pi  \sum\nolimits_t {{\mathbb{E}_{\left( {{{\mathbf{S}}_t},{a_t}} \right) \sim {\rho _\pi }}}\left[ {r\left( {{{\mathbf{S}}_t},{a_t}} \right)} \right]}, 
\end{equation}
where ${\pi ^*}$ represents the optimal policy. Differently, in SAC,  to facilitate the agent to achieve the optimal goals, SAC dynamically optimizes the expected entropy of the policy over ${\rho _\pi }$ during the learning process to flexibly handle  the exploitation and exploration dilemma. Thus, the aim of agent can be reformulated as 
\begin{equation}
    J\left( \pi  \right) = \sum\limits_{t = 0}^T {{\mathbb{E}_{\left( {{{\mathbf{S}}_t},{a_t}} \right) \sim {\rho _\pi }}}\left[ {r\left( {{{\mathbf{S}}_t},{a_t}} \right) + \vartheta \mathcal{H}\left( {\pi \left( { \cdot \left| {{{\mathbf{S}}_t}} \right.} \right)} \right)} \right]}, 
\end{equation}
where $\vartheta$ is a temperature parameter which determines the relative importance of entropy relative to the long-term reward, and $\mathcal{H}\left( {\pi \left( { \cdot \left| {{{\mathbf{S}}_t}} \right.} \right)} \right)$ is the entropy, i.e., $\mathcal{H}\left( {\pi \left( { \cdot \left| {{{\mathbf{S}}_t}} \right.} \right)} \right) =  - {\log _\pi }\left( { \cdot \left| {{{\mathbf{S}}_t}} \right.} \right)$. In addition, SAC follows the classic actor-critic architecture and experience replay mechanism~\cite{haarnoja2018soft, xu2023soft}. The policy ${\pi _\phi }\left( {{a_t}\left| {{{\mathbf{S}}_t}} \right.} \right)$ is generated by the actor network with parameter $\phi$. Meanwhile, to enhance the stability of learning, twin critic networks with the same initial parameter $\theta$  are maintained to avoid overestimation of the soft Q-function ${Q_\theta }\left( {{{\mathbf{S}}_t},{a_t}} \right)$, which is the cumulative expected rewards after taking the action $a_t$ at the state ${{{\mathbf{S}}_t}}$ under the policy $\pi$. 

According to the Bellman equation of the soft Q function, the loss function for the critic network can be expressed by
\begin{equation}
\begin{split}
{\mathcal{L}_Q}&\left( \varphi  \right)  = {\mathbb{E}_{\left( {{{\mathbf{S}}_t},{a_t}} \right) \sim \mathcal{B}}}\left [ {\frac{1}{2}\left( {{Q_\varphi }\left( {{{\mathbf{S}}_t},{a_t}} \right) - \left( {r\left( {{{\mathbf{S}}_t},{a_t}} \right)} \right.} \right.} \right. \\ & \left. {{{\left. {\left. {\frac{{}}{{}} + \gamma {Q_{\bar \varphi }}\left( {{{\mathbf{S}}_{t + 1}},{a_{t + 1}}} \right) - \log {\pi _\phi }\left( {{a_{t + 1}}\left| {{{\mathbf{S}}_{t + 1}}} \right.} \right)} \right)} \right)}^2}} \right],
\end{split}\label{eq:L1}
\end{equation}
where $\mathcal{B}$ represents the replay memory for improving sampling efficiency, and $\bar \varphi$ is the parameter of the target network with soft update, i.e., $\bar \varphi  \leftarrow \tau \varphi  + \left( {1 - \tau } \right)\bar \varphi $ with $\tau  \in \left( {0,1} \right)$. 
Moreover, to maximize the Q value, the loss function for the training of the actor network can be expressed by
\begin{equation}
    {L_\pi }\left( \phi  \right) = {\mathbb{E}_{{{\mathbf{S}}_t} \sim \mathcal{B}}}\left[ {{\mathbb{E}_{{a_t} \sim {\pi _\phi }}}\left[ {\vartheta \log \left( {{\pi _\phi }\left( {{a_t}\left| {{{\mathbf{S}}_t}} \right.} \right)} \right) - {Q_\theta }\left( {{{\mathbf{S}}_t},{a_t}} \right)} \right]} \right]. \label{eq:L2}
\end{equation}
In addition, the adjustment of the temperature parameter can be realized according to the following loss function
\begin{equation}
L\left( \vartheta  \right) = {\mathbb{E}_{{a_t} \sim {\pi _t}}}\left[ { - {\vartheta _t}\log {\pi _t}\left( {{a_t}\left| {{{\mathbf{S}}_t}} \right.} \right) - {\vartheta _t} \mathcal{\bar H}} \right], \label{eq:L3}
\end{equation}
where $\mathcal{\bar H} =  - {\left| \mathcal{A} \right|_{\dim }}$.

We should again highlight that the training and implementation of  K-SAC algorithm is separated. Firstly, an off-line training method is adopted based on  virtual experiences for safeguarding and reducing communication overhead~\cite{xu2023soft,diehl2023uncertainty}. Then, the well-trained model is implemented in real-world scenarios. The optimality, robustness, and learning efficiency are discussed in Section~\ref{sec:simu}. 
Since the predictive frame interpolation at the destination also serves as part of the environment, the  entire  agent-driven semantic sampling algorithm is  outlined in Algorithm~\ref{alg:Agent} after the following elaboration of predictive frame interpolation module.

\begin{figure}[t]
 \centering
\includegraphics[width=1\linewidth]{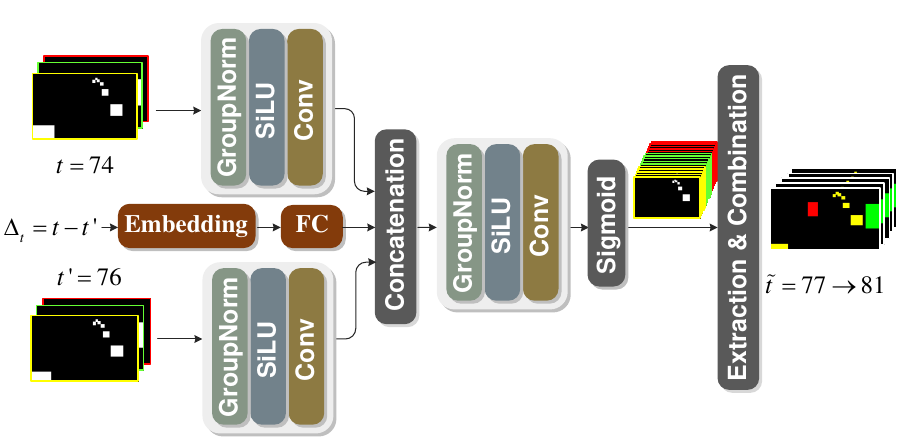}\\
 \caption{Neural network  for predictive frame interpolation.
 }
\label{fig:prediction}
\vspace{-0.5cm}
\end{figure}
\begin{figure}[t]
 \centering
 
 \includegraphics[width = 1\linewidth]{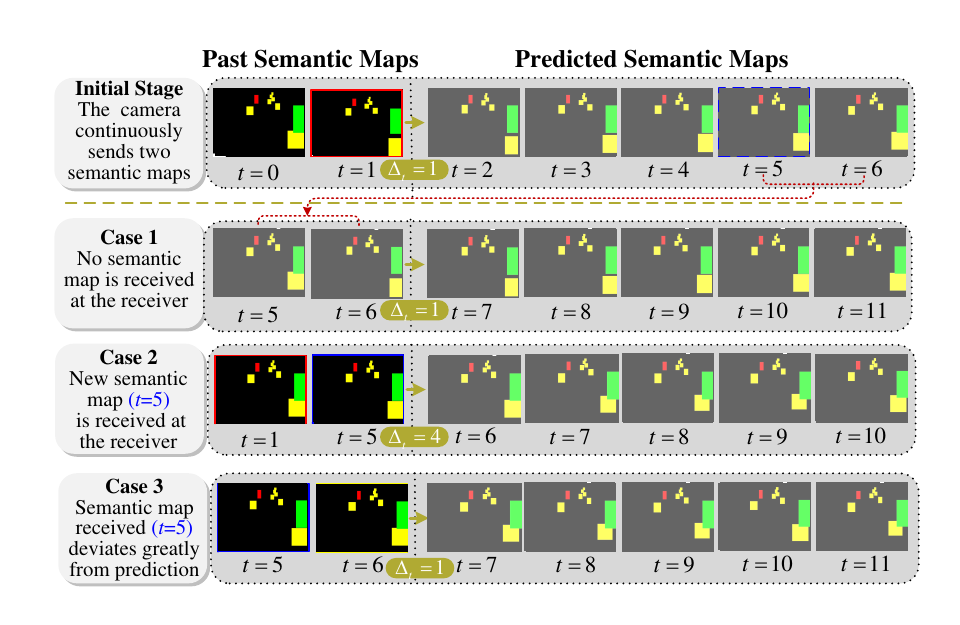}
 
 \caption{Prediction mechanisms at the destination.
 }
\label{fig:drawback}

\vspace{-0.4cm}

\end{figure}

\subsection{Predictive Frame Interpolation Module} \label{SEC:PREDICTION}

The predictive frame interpolation module at the destination plays a crucial role in generating the illusion of real-time updates. When the source does not perform semantic sampling or update semantic information ${{\mathbf{s}}}$ to the destination, the predictive frame interpolation  module  generates predicted visual layout information ${{\mathbf{ {\tilde s}'}}}$ and inputs it into the semantic inference module for timely scene reconstruction. Essentially, this module performs a task similar to video frame prediction but in a much simpler manner.

Specifically, despite significant advancements in DL, the task of video frame prediction continues to present challenges, such as image deblurring, particularly in the long-term prediction~\cite{zhou2020deep}.  Fortunately, since visual layout  exclusively encompasses low-level features and eliminates irrelevant features in the background, the issues caused by the loss of details can be significantly mitigated. Given that  low-level features are widely recognized as being readily captured by  shallow neural networks, we design a lightweight prediction network architecture, as illustrated in Fig.~\ref{fig:prediction}. 
To avoid unnecessary difficulties caused by different vehicle types during the prediction process, the two past visual layouts are first separated into multiple visual layouts, each containing only one type of vehicle. These layouts are then fed into two ResBlocks with shared parameters to extract features using the batching technique.  
Simultaneously, given that the intervals for semantic sampling exhibit variability, we also take  the embedding of the temporal discrepancy $\Delta_t$ between the timestamps of two visual layouts  as  the  inputs,
which is then amalgamated with  both derived feature maps corresponding to the two past visual layouts. The fused feature is subsequently processed by another ResBlock to predict the visual layout for each type of vehicles.   
{Finally, through a extraction and combination module, the visual layout for a scene can be output and used as the prompts to generate the remote scene, if required.}
To reduce the computational overhead associated with frequent predictions, we set that the module predicts next $P$ visual layouts at once. 
{\color{Reds} In this sense, a larger value of $P$ means lower system complexity. However, increasing $P$ also raises the computational power requirement at the destination and amplifies the impact of vehicle speed variations on prediction accuracy, posing greater challenges to maintaining prediction accuracy. Therefore, the selection of $P$ requires a trade-off between these factors.}

To align with adaptive semantic sampling at the source, the visual layout prediction mechanisms at the destination for different cases are as follows.

\textbf{Case~1}: At the initial STI, the source continuously samples and transmits the semantic information ${{\mathbf{s}}_0}$ and  ${{\mathbf{s}}_1}$ corresponding to the scenes within two consecutive STIs. Subsequently, the semantic inference module first transforms semantic information into the visual layout ${{{\mathbf{\tilde s}}}_{00}}$ and ${{{\mathbf{\tilde s}}}_{01}}$. Then, the predictive frame interpolation module predicts the next $P$ visual layouts ${{{\mathbf{\tilde s}'}}_1}$ ... ${{{\mathbf{\tilde s}'}}_{P}}$ based on ${{{\mathbf{\tilde s}}}_{00}}$ and ${{{\mathbf{\tilde s}}}_{01}}$.

\textbf{Case 2}: In the following STI $t$, if no updated semantic information is received and the visual layout for that moment has already been predicted, the predictive frame interpolation module feeds ${\mathbf{\tilde s}}{'_t}$ directly to semantic  inference module. 
Conversely, if ${\mathbf{\tilde s}}{'_t}$ falls outside the scope of the previous prediction, the  prediction module uses ${\mathbf{\tilde s}}{'_{t-1}}$,  ${\mathbf{\tilde s}}{'_{t-2}}$ and ${\Delta _t} = 1$ as the inputs for the next round prediction. Notably, the latter scenario indicates that the remote scene undergoes minimal changes, ensuring that successive predictions do not result in much cumulative deviation.

\textbf{Case 3}: In the following STI $t$, if new layout information ${{\mathbf{s}}_t}$ is received at the destination,  the  prediction module first compares the visual layout ${{{\mathbf{\tilde s}}}_t}$ transformed from ${{\mathbf{s}}_t}$ with the predicted visual layout ${{{\mathbf{\tilde s}}'}_t}$. If the prediction deviation $D_t$ falls below a predetermined threshold, the prediction module promptly conducts the next round of prediction based on the updated ${{{\mathbf{\tilde s}}}_t}$ and the  last sampled ${{{\mathbf{\tilde s}}}_{\hat t}}$. This approach minimizes the spread of prediction deviations. In contrast, if $D_n$ exceeds the preset threshold, it indicates a sudden change in the remote scene. In such a case, the source is informed to continue performing the sampling at STI $(t+1)$ and initiate a new prediction process at the destination.
  
It is important to note that the above mechanisms need to be integrated into the dynamics of the environment that semantic sampling agent interacts with. Furthermore, given that the latter scenario in Case 3 is an extreme instance that can serve as a starting point for a new episode, it is not treated as a distinct entity in environmental modeling. The full realization of the semantic sampling in the proposed SemCom framework has been highlighted in Algorithm~\ref{alg:Agent}.


\begin{table}[]
\footnotesize
 \centering
 \caption{\color{Reds}Simulation Parameter Settings}
\renewcommand{\arraystretch}{1.1}
\begin{tabular}{m{3.65cm}|m{0.5cm}|m{2.6cm}|m{0.4cm}}
\hline \hline
\multicolumn{4}{c}{\textbf{Wireless Network Parameter Settings}}   \\  \hline
\multicolumn{2}{l|}{Small-scale fading model} & \multicolumn{2}{l}{$\mathcal{F}$ composite fading mode~\cite{8638956}} \\ \hline                        \multicolumn{2}{l|}{Pass loss (average channel gain) [dB]} & \multicolumn{2}{l}{$35.3 + 37.6{\log _{10}}\left( d \right)$~\cite{liu2021accelerating}}                                               \\ \hline
Fading severity  $m$                        & 6 & Bandwidth $W$ [kHz]     & {1}    \\ \hline
Wireless Distance $d$ [m]        & 100     &SNR
threshold  [dB]               &    15  \\  \hline              
Noise power [dBm/Hz]      & -90 & Shadowing shape  $m_s$      & {6}    
\\ \hline \hline
\multicolumn{4}{c}{\textbf{Hyperparameters for Generative Semantic Inference Module}}                                                                         \\ \hline 
{Learning rate}                           & {2e-5} & {Batch size}         & {1}    \\ \hline
Denoising step  $N$                                              & 1000                      & Guidance scale  $k$                             & 8                         \\ \hline \hline
\multicolumn{4}{c}{\textbf{Hyperparameters for Predictive Frame Interpolation Module}}                                                                        \\ \hline 
{Learning rate}                           & {2e-5} & {Batch size}         & {5}    \\ \hline \hline
\multicolumn{4}{c}{\textbf{Hyperparameters for K-SAC based Semantic Sampling Agent}}                                                                          \\ \hline 
{Memory Capacity}                         & {1e5}  & {Batch size}         & {1024} \\ \hline
{Learning rate for policy network}        & {1e-5} & {Soft Update}        & {0.2}  \\ \hline
{Learning rate for critic network}        & {2e-5} & {Observation window} & {150}  \\ \hline
\multicolumn{2}{l|}{Learning rate for temperature parameter} & \multicolumn{2}{l}{1e-5}    \\ \hline
\end{tabular}
\label{tbl: simu}
\end{table}

\begin{figure}[t]
 \centering

\includegraphics[width=1\linewidth]{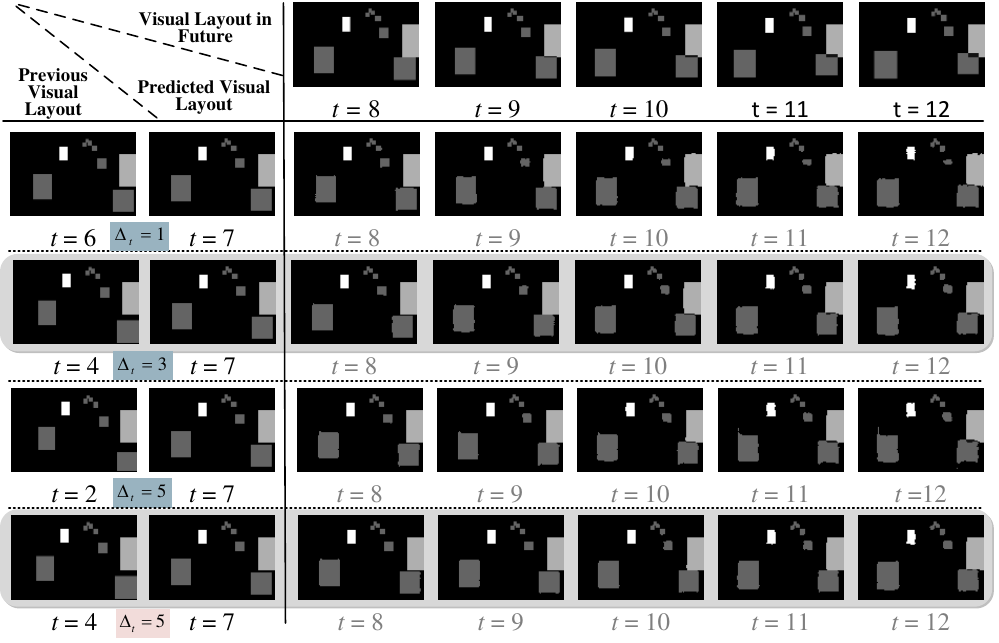}

 \caption{Visual results for visual layout prediction. 
 }
\label{fig:FPprediction}
\end{figure}
\begin{figure*}[t]
 \centering
\includegraphics[width=0.9\linewidth]{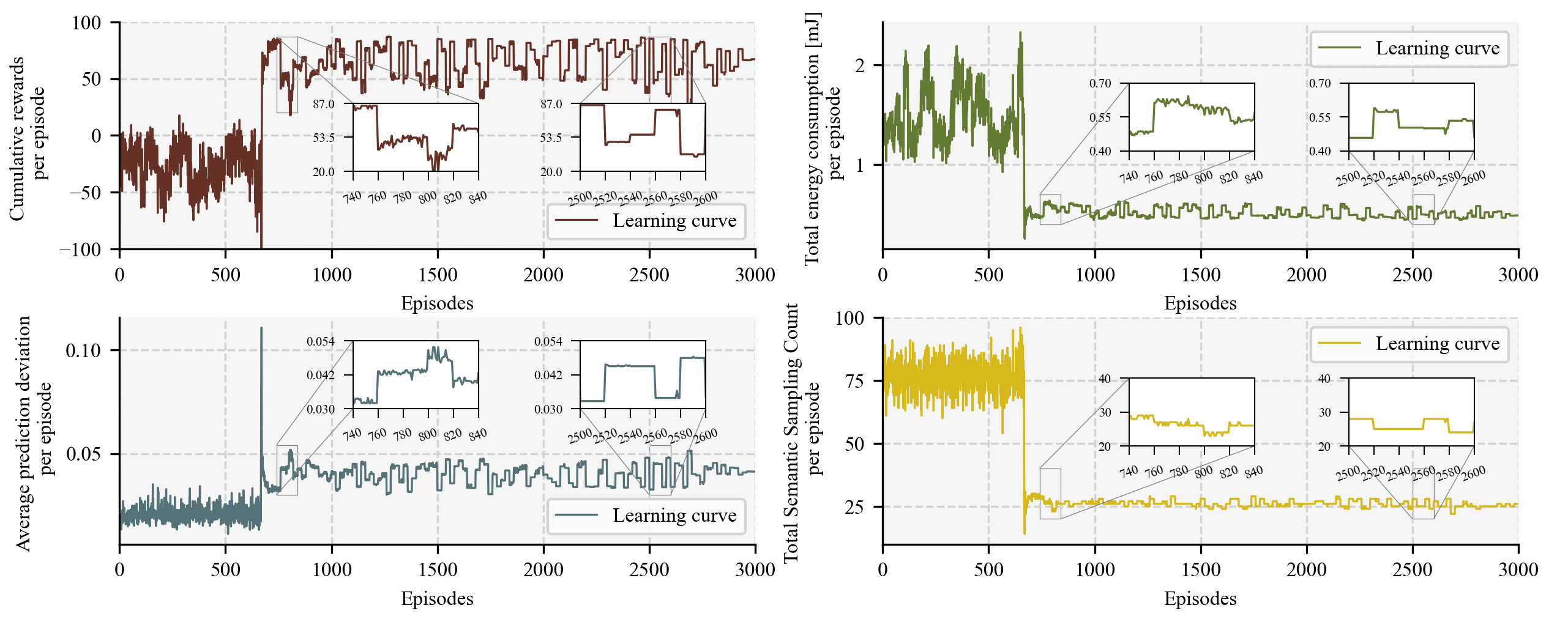}

\caption{Learning performance for K-SAC agents with \textit{random initial scene with different traffic pattern} in terms of cumulative reward, total energy consumption, average prediction deviation, and total semantic sampling count.
 }
\label{fig:RLlearnincurve}
\vspace{-0.4cm}
\end{figure*}
\begin{figure*}[t]
 \centering

\includegraphics[width=0.9\linewidth]{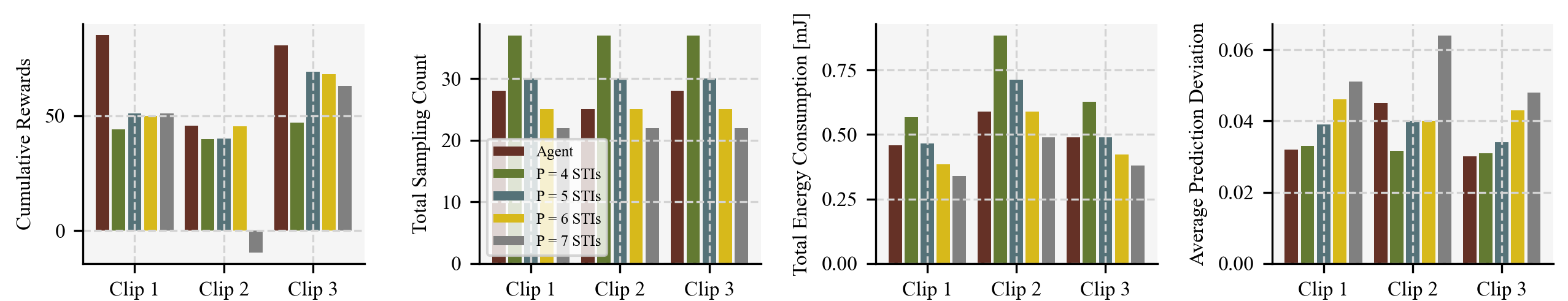}

 \caption{Performance comparison about {\color{Reds}A-GSC where semantic sampling is performed adaptively by the agent, and {\color{Reds}GSCs} where the semantic sampling is performed periodically with multiple periods (denoted by $P$)} for different footage clips. {\color{Reds}Semantic extraction in all GSC paradigms is performed by the semantic extraction module stated in Section~\ref{sec: extraction}}.
 }
\label{fig:TesT}
\vspace{-0.3cm}
\end{figure*}
\section{Simulation}
\label{sec:simu}
\subsection{Simulation Setup}
\subsubsection{Datasets}
We select typical bidirectional road traffic scenarios from  UA-DETRAC dataset~\cite{wen2020ua}  to serve as the training data for the generative semantic inference module, predictive frame interpolation, as well as the semantic sampling agent.
Specifically, as this work excludes the optimization of the object detection module, we directly use  the information about bounding box and the type of all vehicle objects recorded in the \textit{Annotations} as the extracted textual semantic information. Moreover, when mapping the visual layout, we set the pixel values corresponding to the four categories of \textit{car}, \textit{bus}, \textit{van}, and \textit{others} to 1, 2, 3, and 4, respectively. 
Moreover, during the training, the image size is reshaped into $\left( {240, 160} \right)$. {\color{Reds}Meanwhile,  for the wireless transmission parameter settings, the detailed parameter settings can be found in Table~\ref{tbl: simu}.}

\subsubsection{{\color{Reds}Network design and} hyperparameters} 
For neural network of the predictive frame interpolation
module shown in Fig.~\ref{fig:prediction}, the number of channels in each ResidualBlock is set to 8, and the number of the channels in the output layer is set to 5, i.e., $P=5$. The dimension of the time embedding is 
set to 32.
For the generative semantic inference module as shown in Fig.~\ref{fig:Unet}, it comprises a number of channels equal to [64, 64, 128, 128, 256, 256, 512, and 512] and with a linear variance schedule. Meanwhile, the dimension of the time embedding is 
set to 256. 
Finally, we adopt the classifier-free guidance ~\cite{wang2022semantic} with the guidance scale of $k=8$.   For the K-SAC-based semantic sampling policy, both the policy network and the two critic networks are designed with three layers, with 300, 200, and 200, respectively.   In addition, for all the models, the Adam optimizer is adopted with betas=(0.9, 0.999). {\color{Reds}Moreover, more hyperparameters during the training for each model have been summarized in Table~\ref{tbl: simu}.}

\subsection{Performance Analysis}
We first analyze the performance of each module sequentially according to their operational order. Finally, we present a comprehensive end-to-end performance analysis.
\subsubsection{Predictive frame interpolation} 
\label{sec:s1}
The predicted visual layouts with different sampling intervals are shown in Fig.~\ref{fig:FPprediction}. Notably, during predictions with ${\Delta _t} = 1$ STI, ${\Delta _t} = 3$ STI, and ${\Delta _t} = 5$ STI, the positional consistency of objects is preserved. This ensures the fidelity of semantic information between the real-world scene and the generated scene. 
However, since the module can only grasp the relative positions of the objects in the two layouts and the time differences based on the inputs, the prediction relies on an implicit average driving speed. As a result, due to the non-uniform speed, careful observation of Fig.~\ref{fig:FPprediction} reveals that there are some slight discrepancies corresponding to different sampling time intervals. This can be further confirmed by comparing the second and fourth rows in Fig.~\ref{fig:FPprediction}, which use the same layouts for prediction but different sampling intervals. Meanwhile, this observation also  confirms the validity of the prediction module. {\color{Reds}In addition, as shown in Fig.~\ref{fig:FPprediction}, the contours of the later predicted frames become increasingly blurry, which may also interfere with the evaluation of semantic accuracy. In this sense,  even if computational power permits, the value of $P$ should not be too large. Therefore, considering both semantic accuracy and the complexity of frequent predictions, we select $P=5$ based on experimental results.}
\begin{figure*}[t]
 \centering

\includegraphics[width=0.9\linewidth]{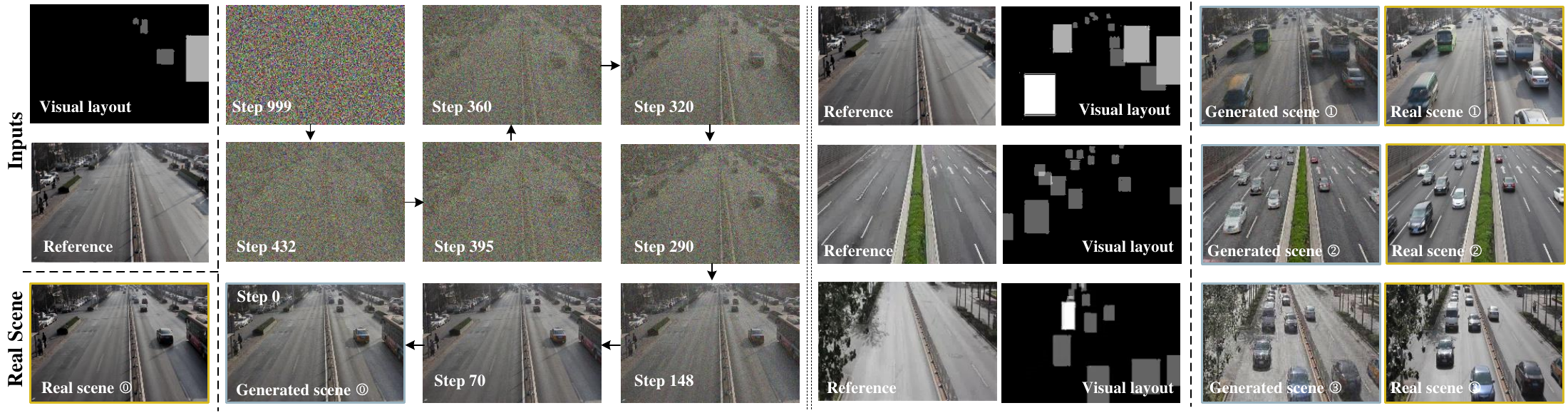}
 \caption{Visual results for generative semantic inference.  
 }
\label{fig:compare}
\vspace{-0.5cm}
\end{figure*}

\subsubsection{K-SAC based semantic sampling}
To enhance the robustness and adaptability of the algorithm, we select 500 frames of continuous surveillance scenes as the training environment. These frames include scenarios with varying vehicle types, vehicle counts, vehicle speeds, and changes in vehicle speeds. Specifically, the real-world semantic information is transformed into visual layouts based on the annotations from UA-DETRAC XML files. The visual layout prediction utilizes the model trained in Section~\ref{sec:s1}.  The convergence of the algorithm is illustrated in Fig.~\ref{fig:RLlearnincurve}, where an episode refers to an observation window. 

It is important to note that during the training process, \textit{the initial state (the scene with different traffic pattern) is randomly regenerated every 20 episodes}. This differs from the majority of current RL training methodologies, which use the same or similar initial states across different episodes, allowing the reward function to converge to a stable value for each episode. In our training process, the episodes with different initial states always encompass different observed environments, where variations in the number of vehicles and the vehicular speed  inevitably affect the communication system's performance in terms of energy consumption and prediction accuracy. Due to the inherent randomness of traffic flow, it is neither feasible nor necessary to compensate for its impact on performance through the design of the reward function. Consequently, as expected, the cumulative reward  for different observation windows deviates after convergence of the algorithm. Nevertheless, aside from the overall increase in the cumulative reward function value indicating the  convergence of the algorithm, this stability of the cumulative reward for the same observed environment, as locally magnified in Fig.~\ref{fig:RLlearnincurve}, also corroborates the  convergence of the algorithm. From Fig.~\ref{fig:RLlearnincurve}, we can observe that during the initial convergence phase, from episode 740 to episode 840, all the performance metrics exhibit differences across the five  observation windows. However, there are still noticeable fluctuations within each observation window. In contrast, during the late convergence phase, from episode 2500 to episode 2600, the performance metrics within the different windows have stabilized, with only minor variations attributable to  inherent randomness of the strategy. 
Moreover, by comparing the locally magnified view of the total energy consumption with the locally magnified view of the number of samples, we can observe that fewer samples often correspond to higher energy consumption. This reflects the agent's energy-saving awareness and adaptive capability. 

Furthermore, the superiority of agent-assisted semantic sampling compared to existing periodic sampling strategies  has been demonstrated in Fig.~\ref{fig:TesT}. Specifically,   we consider periodic sampling strategies with periods of 4, 5, 6, and 7 STI for three footage clips with different initial moments\footnote{The videos about the three footage clips can be found at \textbf{\url{https://youtu.be/pJZzDnNspAc}}}. 
.Firstly, from the perspective of the comprehensive evaluation indicator, cumulative reward, all sampling strategies exhibit varying performances across different footage clips. Among them, the agent-assisted sampling strategy, due to its robustness and adaptability, achieved significantly superior performance in all three footage clips. Additionally, there is no optimal sampling period for the periodic sampling strategies because of the dynamic changes in the environment. For example,  the periodic sampling strategy with a period of 7 STI achieved the best performance among the four periodic sampling strategies in  Clip 1, whereas in Clip 2, its performance was the worst, resulting in negative values.
The reason for this can be understood by comparing the subsequent three subplots. Although the periodic sampling strategy has the same sampling counts in different footage clips, it achieves the lowest energy consumption in Clip~1. This implies that the smaller number of vehicles in Clip~1 allows for more sampling with the same energy consumption, further reducing prediction deviation and improving overall performance. The agent-assisted sampling strategy exactly achieves this effectively. In Clip~2, the prediction deviation of the periodic sampling strategy with a period of 7~STI significantly increases compared to the strategies with periods of 5 and 6 STI, suggesting that there may be sudden changes in traffic flow within a certain sampling interval. The agent-assisted sampling strategy addresses this issue by increasing a few sampling amount with minimal energy consumption, further demonstrating its robustness and adaptability. {\color{Reds}From the overall perspective of Fig.~\ref{fig:TesT}, the average sampling interval for the agent-assisted semantic sampling is approximately between 5 STI and 6 STI for different footage clips, which is close to the value of $P$. This is because, although a corresponding prediction mechanism at the destination side compensates for the semantic information loss caused by sampling, as mentioned in Section~\ref{sec:s1}, the accumulation of prediction errors increases with longer prediction interval. Especially when the next round of prediction is based on two consecutive predicted frames, this error is further amplified.}

\begin{figure}[t]
 \centering
 
 \includegraphics[width = 0.9\linewidth]{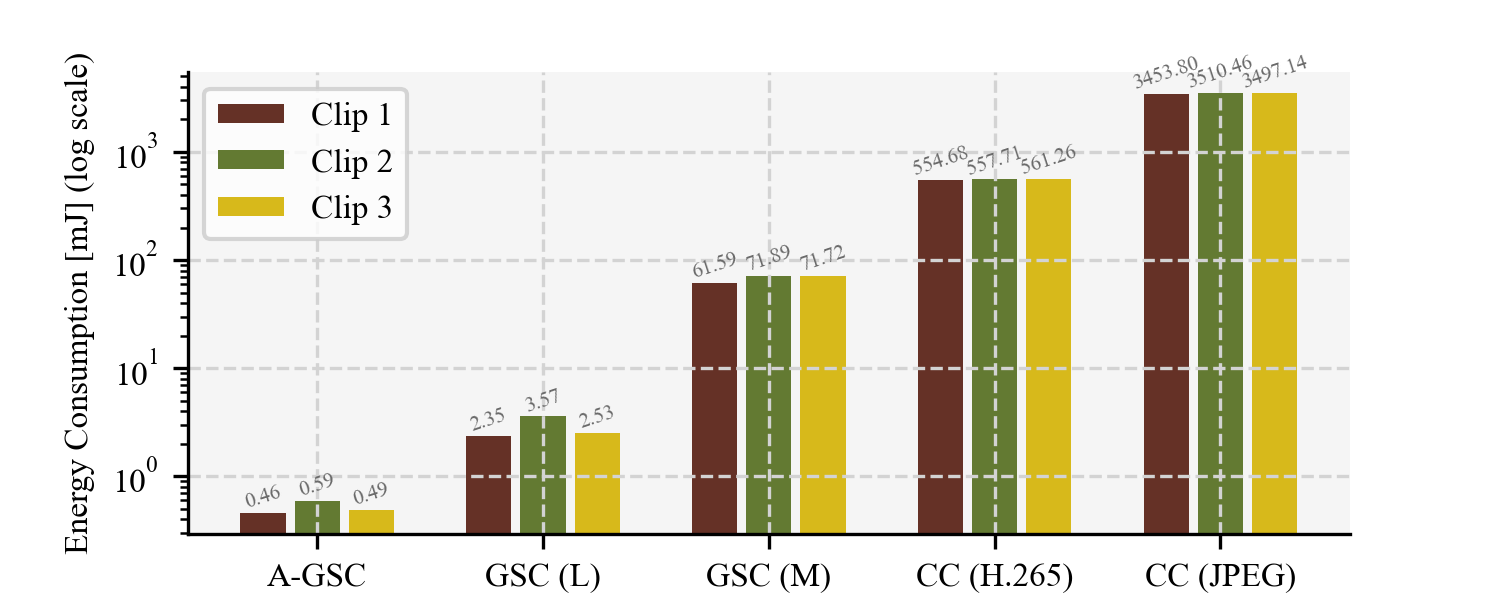}

 \caption{Comparison of energy consumption. 
 }
\label{fig:energy}

\vspace{-0.3cm}

\end{figure}

\begin{figure}[t]
 \centering
 
 \includegraphics[width = 0.9\linewidth]{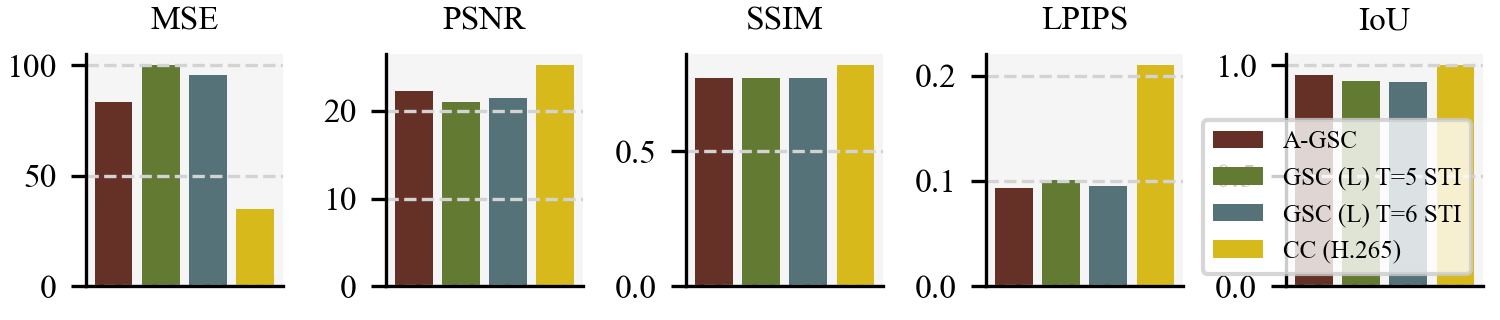}
 
 \caption{Comparison of the accuracy of the scene reconstruction with different strategies under multiple metrics.
 }
\label{fig:metric}

\vspace{-0.5cm}

\end{figure}

\begin{figure}[t]
 \centering
 
 \includegraphics[width = 1\linewidth]{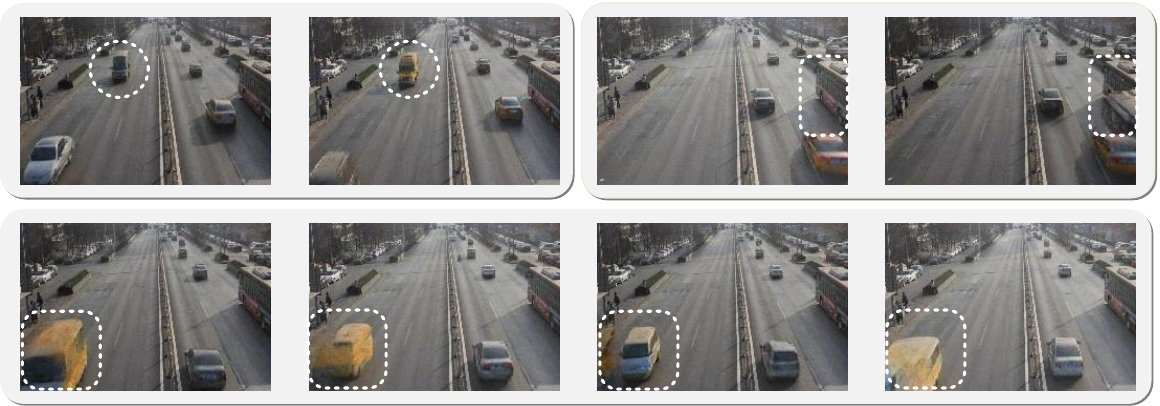}
 
 \caption{Limitations of proposed A-GSC.
 }
\label{fig:drawback}

\vspace{-0.4cm}

\end{figure}
\subsubsection{Generative semantic inference}
The  visual results of the generative semantic inference can be found in Fig.~\ref{fig:compare}. Specifically, the left side of the figure shows the specific denoising process, and the right side shows more examples generated under different references.  By comparing the generated image and the real image, it becomes evident that, while there exist differences in the background details and coloration of the vehicles, the positional alignment of the vehicles remains accurate. This achievement signifies the system's proficiency in preserving essential semantic information. However, it should be noted that if the boundary vehicle fails to be detected at the source, the generated image at the destination will not contain that information (as shown in Example \ding{172} in Fig.~\ref{fig:compare}
).

\subsubsection{Overall performance} 
Finally, we analyze the overall performance gain of the proposed A-GSC framework. To demonstrate the superior performance in terms of energy efficiency,  we conduct a comparison of the proposed A-GSC between the existing generative SemComs (GSCs) with semantic map and textual layout as prompt (denoted by GSC (M) and GSC (L) respectively), as well as  conventional communications (CCs) with the source coding H.265\footnote{We employ FFmpeg, leveraging its integration with the libx265 library for H.265/HEVC encoding, and acquire the data size of each encoded frame.} and JPEG, respectively. 
Although the advanced H.265 encodes only the differences between blocks of pixels within a frame or between consecutive frames, it treats all changes equally. Therefore, due to the varying dynamics of objects and backgrounds, H.265 still requires {\color{Reds}10 times more}   energy consumption compared to GSC (M), as shown in Fig.~\ref{fig:energy}. Furthermore, from the comparison of GSC (L) and GSC (M), We can observe that the introduction of image-to-text conversion is well worth it in terms of improving energy efficiency, {\color{Reds}as it can reduce the amount of transmitted data to approximately 1/30 of the original}. Moreover, comparing the energy consumption of A-GSC and the existing GSC, the integration of semantic sampling agent {\color{Reds}further reduces energy consumption to 1/5 from the temporal domain, compared to GSC (L).}  Considering the substantial differences in energy consumption across different communication paradigms, \textit{log scale} is adopted in Fig.~\ref{fig:energy}.     {\color{Reds}Overall, as can be clearly seen from Fig.~\ref{fig:energy}, the proposed A-GSC reduces transmission energy consumption by approximately 1,000 times compared to the current H.265-based encoding.  }

After demonstrating the great potential for energy savings, we evaluate the performance of scene reconstruction based on multiple metrics. We compare our method with GSC (L) periodic sampling at intervals of T=5 STI and T=6 STI, which have shown better performance as depicted in Fig.~\ref{fig:TesT},  as well as with conventional coding (CC) using H.265. Moreover, it is noted that in the CC with H.265, we resize the frame into (120, 80) before transmission. Then, during the reconstruction, we resize it to (240, 160). This is consistent with the use of the (120, 80) frame size in the visual layout prediction.
The visual video for the source data sensed at the source and the reconstructed scene by multiple communication paradigms   can be found at \textbf{\url{https://youtu.be/9tNKq0RQ58Y}}.
From the video, we can see that due to the stochastic nature of the inference process, the color of the vehicles may change across different scenes, and the brightness of the image also fluctuates a bit.  Both  observations restrict GSCs' performance in the evaluation under the conventional metrics of MSE and PSNR, the SSIM, but they are still at a satisfactory level as shown in Fig.~\ref{fig:metric}. In contrast, under the semantic metric Learned Perceptual Image Patch Similarity (LPIPS) and Intersection over Union (IoU) about the considered task, the proposed algorithm shows its superiority. {\color{Reds}Compared to periodic sampling, the performance of agent-assisted GSC improves by approximately 6\% and 4\%}. In addition, due to the photo-realistic generation capability of the diffusion model and high compression rate adopted in H.265, the GSCs' performance {\color{Reds}is 2 times that of}  conventional coding (CC) with H.265 under the LPIPS metric.  
Overall, the above demonstrates that A-GSC framework can also achieve fulfilling performance.

However, the framework currently has limitations. Fig.~\ref{fig:drawback}
 highlights three potential unexpected issues. First, while the accuracy of vehicle position changes can be maintained in a video, the color of a vehicle may  be altered. Second, for larger objects such as buses, it is occasionally hard to determine whether a single vehicle or two overlapping vehicles are present. Third, with boundary objects, what should be relatively large incomplete objects may sometimes be interpreted as a smaller complete vehicle. Although these errors do not affect semantic accuracy, further improvements are necessary.

{\color{Reds}\subsection{Complexity Analysis of Training and Deployment}
In the proposed A-GSC framework, multiple trainable modules are incorporated, including the semantic extraction module, semantic sampling agent, predictive frame interpolation module, and generative semantic inference module. Nevertheless, this explainable modular design allows each module to be trained independently, significantly reducing the overall training complexity. Specifically, for the source end, a well-trained object detection algorithm can be utilized as a pre-trained model, which greatly mitigates the complexity of training from scratch. Additionally, for the semantic sampling agent, since it performs a binary decision task and the state size is relatively small, the convergence rate is fast, requiring only a few hundred episodes to reach a satisfactory performance, as shown in Fig.~\ref{fig:RLlearnincurve}). Moreover,  due to the small dimensions of the action and state spaces, the storage space required for an experience replay buffer of 100,000 pieces of experience is only 8.5 MB. Therefore, it also has minimal requirement for storage.
On the destination side, the visualized layout information primarily consists of low-level features, significantly reducing both model design and training complexity compared to existing video frame prediction methods, with only 500 episodes needed. Furthermore, for the generative semantic inference module, as this task does not involve complex cross-modal feature alignment like stable diffusion with text-based prompts, its model size and training difficulty are greatly reduced compared to general generative models such as stable diffusion, requiring only 400 episodes.
Meanwhile, if this communication framework is to be applied to other scenarios with richer and more complex semantic information, well-trained models like ControlNet can be employed as pre-trained models for further fine-tuning. Overall, despite the increase in the number of modules requiring training, the training complexity is much lower compared to JSCC  SemCom approaches based on Transformers, which typically require millions or even tens of millions of episodes~\cite{yang2024swinjscc}.

Additionally, for the deployment process, a lightweight YOLO object detection algorithm designed for mobile devices can be used as the semantic extraction module. Furthermore, the semantic sampling agent, with a small-scale actor network consisting of 0.102018 million parameters, requires only 0.102656 million FLOPs per inference. Compared to the lightweight and widely used YOLOX model~\cite{ge2021yolox}, which requires 1.08G FLOPs and 0.91 million parameters, we believe that integrating the K-SAC agent into embedded devices is feasible. Meanwhile, on the destination side, the long-term frame prediction in the predictive frame interpolation module has reduced the complexity of frequent predictions. For the generative semantic inference module, due to its smaller model size, the inference latency for a single denoising step is also reduced. Furthermore, it is worth highlighting that advanced diffusion models such as InstaFlow and DDIM have optimized the inference process of classic DDPMs by reducing the number of denoising steps, thereby lowering the inference latency to 0.1 seconds and 1 second, respectively. This demonstrates the potential feasibility of applying diffusion models in surveillance systems. 
}
\section{Conclusion}
\label{sec:conclusion}

In this paper, we have focused on a remote surveillance scenario. In contrast to the existing research efforts for SemCom, which have focused solely on semantic extraction or semantic sampling, we have seamlessly integrated both together by introducing the GAI and an RL-based agent into SemCom system design. Specifically, in the proposed A-GSC framework, the semantic information has served as an explainable prompt of textual layout information transmitted to the semantic decoder while assisting in semantic sampling. To consistently display the remote scene, we have integrated prediction capabilities into the decoder design, to complement the flexible semantic sampling at the semantic encoder. Our simulations have demonstrated notable improvements in both energy efficiency and reconstruction accuracy. In the future, we will further extend the work to a multi-sensor scenario, and aim to jointly optimize semantic compression as well as semantic scheduling.









\bibliographystyle{IEEEtran}
\bibliography{ref}



\end{document}